\documentclass[10pt,journal,compsoc]{IEEEtran}
\usepackage{graphicx}
\usepackage{amsmath,amssymb} 
\usepackage{color}
\usepackage{todonotes}

\usepackage{algorithm}
\usepackage{algpseudocode}  
\usepackage{multirow}
\usepackage{makecell}
\usepackage{booktabs}
\usepackage{subfigure}
\usepackage{hyperref}
\usepackage{url}
\usepackage{ulem}
\usepackage{tabularx,booktabs}
\usepackage{arydshln}

\ifCLASSOPTIONcompsoc
  \usepackage[nocompress]{cite}
\else
  \usepackage{cite}
\fi

\ifCLASSINFOpdf
\else
\fi

\begin{document}

%
\title{Automatic Depression Assessment using Machine Learning: A Comprehensive Survey}
%
%
%
%
\author{Siyang Song,~\IEEEmembership{}
        Yupeng Huo,~\IEEEmembership{}
        Shiqing Tang,~\IEEEmembership{}
        Jiaee Cheong,~\IEEEmembership{}
        Rui Gao,~\IEEEmembership{}
        Michel Valstar~\IEEEmembership{}
        and Hatice Gunes~\IEEEmembership{}

\IEEEcompsocitemizethanks{

\IEEEcompsocthanksitem Siyang Song, Yupeng Huo and Shiqing Tang are with HBUG Lab, Department of Computer Science, University of Exeter, Exeter, United Kingdom. Corresponding Author: Siyang Song (s.song@exeter.ac.uk)

\IEEEcompsocthanksitem Jiaee Cheong and Hatice Gunes are with the AFAR Lab, Department of Computer Science and Technology, University of Cambridge, Cambridge, United Kingdom.

\IEEEcompsocthanksitem Rui Gao is University of California, Los Angeles, United States.

\IEEEcompsocthanksitem  Michel Valstar is with the Computer Vision Lab, School of Computer Science, University of Nottingham, Nottingham, United Kingdom.}

\thanks{Manuscript received June 22, 2025}
}

%
%

\markboth{IEEE TRANSACTIONS}%
{Shell \MakeLowercase{\textit{et al.}}: Bare Advanced Demo of IEEEtran.cls for IEEE Computer Society Journals}

\IEEEtitleabstractindextext{%
\begin{abstract}

Depression is a common mental illness across current human society. Traditional depression assessment relying on inventories and interviews with psychologists frequently suffer from subjective diagnosis results, slow and expensive diagnosis process as well as lack of human resources. Since there is a solid evidence that depression is reflected by various human internal brain activities and external expressive behaviours, early traditional machine learning (ML) and advanced deep learning (DL) models have been widely explored for human behaviour-based automatic depression assessment (ADA) since 2012. However, recent ADA surveys typically only focus on a limited number of human behaviour modalities. Despite being used as a theoretical basis for developing ADA approaches, existing ADA surveys lack a comprehensive review and summary of multi-modal depression-related human behaviours. To bridge this gap, this paper specifically summarises depression-related human behaviours across a range of modalities (e.g. the human brain,  verbal language and non-verbal audio/facial/body behaviours). We focus on conducting an up-to-date and comprehensive survey of ML-based ADA approaches for learning depression cues from these behaviours as well as discussing and comparing their distinctive features and limitations. In addition, we also review existing ADA competitions and datasets, identify and discuss the main challenges and opportunities to provide further research directions for future ADA researchers.

\end{abstract}

}

\maketitle

\IEEEdisplaynontitleabstractindextext

%
\IEEEpeerreviewmaketitle

\section{Introduction}
\label{sec:introduction}

Depression, also called major depressive disorder (MDD), is a significant mental disorder that negatively impacts around 3.8\% of the world population, ranging from mild to severe conditions with hallucinations and delusions,  
i.e., with one in 15 adults diagnosed with depression every year \cite{thapar2022depression}, which would physiologically, mentally and socially impair individuals, including damaging brain structures \cite{van2013paralimbic,lu2012impaired} and perception ability \cite{bourke2010processing}, causing chronic physiological illness \cite{gan2014depression}, as well as substantial financial costs \cite{thapar2022depression,lepine2011increasing}. Depression can be categorised into several types, including clinical depression (major depressive disorder, MDD), persistent depressive disorder (PDD), disruptive mood dysregulation disorder (DMDD), premenstrual dysphoric disorder (PMDD) and depressive disorder triggered by other medical conditions \footnote{\url{https://my.clevelandclinic.org/health/diseases/9290-depression}}. The predominant characteristics of depression are the depressed mood or dysphoria reflected by various daily symptoms such as diminished interest in daily activities, alterations in sleep and appetite, feelings of guilt and hopelessness, fatigue, difficulties in concentration, and even suicide attempts \cite{abbey1991chronic,paykel1977depression,benca2008insomnia}.

As a result, objective and accurate depression screening is crucial for raising early assessment and intervention. Existing clinical depression assessments typically rely on structured or semi-structured interviews conducted by professional psychologists \cite{smith2013diagnosis}, where specifically-designed multi-point questionnaires (e.g., Hamilton Rating Scale for Depression (HAM-D) \cite{hamilton1960rating} and PRIME-MD Patient Health Questionnaire (PHQ) \cite{kroenke2001phq}) are required to be filled by either patients or psychologists to describe depression symptoms \cite{fletcher2008adolescent}. However, these strategies frequently suffer from: (i) substantial costs charged by psychologists and clinical equipment usage; (ii) long assessment period caused by complex protocols (e.g. psychologist appointment and traveling); (iii) subjective assessment outcomes caused by disparities in psychologists' expertise; and (iv) biased outcomes as patients may exaggerate/deemphasise crucial facts or symptoms in verbal reports/questionnaires. Such limitations can cause delays and unreliability in depression assessment, prolonging depression patients' suffering. 

As a result, there is a recognised necessity within the medical community to explore cheap, fast, objective (repeatable) and reliable solutions for depression assessment, e.g., incorporating machine learning (ML) techniques for automatic depression assessment (ADA).
These ADA approaches are generally built on consistent biological and physiological findings that various human internal and external behaviours are informative for reflecting depression status in clinical trails. Specifically, multiple brain behaviours such as abnormal grey matter changes \cite{van2013paralimbic}, 
physiological behaviours (e.g., `accelerated aging' \cite{wolkowitz2010depression} and body shape changes  \cite{milaneschi2019depression,fu2023shared}), 
as well as external expressive behaviours including facial behaviours (e.g., reduced expressions \cite{gupta2009major}), 
gestures (e.g., fewer gestures \cite{hinchliffe1975study}), body movements \cite{deligianni2019emotions,gimeno2024reading}, verbal language behaviours (e.g., frequent first person singular usage \cite{nilsonne1988measuring,pennebaker2001linguistic}), and non-verbal vocal behaviours (e.g., 
reduced pitch range \cite{hollien1980vocal}), can indicate or be associated with depression. 
%



The majority of existing ADA solutions infer depression from easily accessible human expressive behaviours (e.g., audio, facial and language behaviours), including traditional ML-based solutions \cite{kaya2014ensemble,pampouchidou2016depression,muzammel2021end} that combine hand-crafted audio, visual or audio-visual features (e.g., Local Binary Pattern (LBP) and Mel-frequency cepstral coefficients (MFCCs)) with traditional ML regressors/classifiers (e.g., Support Vector Machine Regression (SVR), Random Forest (RF), Logistic Regression (LR) and k-nearest neighbor (KNN)), and DL-based ADA approaches \cite{jung2024hique,xu2024two,niu2020multimodal,khowaja2024depression,ye2024dep,fang2023multimodal} leveraging Convolution Neural Networks (CNNs), Transformers or Graph Neural Networks (GNNs) for end-to-end depression feature extraction and assessment. 
To address ethical and privacy issues, in some occasions only de-identified behaviour data are provided, and thus some ADA approaches \cite{gahalawat2023explainable,abbasi2022statistical,pan2023integrating,gong2017topic,rodrigues2019multimodal,wang2024facialpulse} were specifically developed on anonymous facial behaviour primitives (e.g., facial Action Units (AUs) and head poses) and de-identified audio/text signals. Among existing expressive behaviour-based ADA approaches, majority assess depression status based on an entire given clip (usually less than 30 minutes), while only a few studies \cite{girard2013social,ahmad2021cnn,shang2021lqgdnet} infer depression from static face images as long-term behaviours (typically lasting several weeks) may be more reliable indicators of depression \cite{song2022spectral,de2021mdn}. Alternatively, other ADA solutions were  developed for analysing brain behaviours recorded by non-invasive electroencephalography (EEG) \cite{zhang2024novelEEG,shen2023depression} or magnetic resonance imaging (MRI) \cite{chen2021convolutional,gao2023classification,mousavian2021depression}. Similarly to external expressive behaviour-based ADA approaches, two types of pipelines have been explored: (i) hand-crafted pipeline that applies traditional ML predictors to process manually defined features (e.g., spectral EEG features reflecting depression-related human brain electrical activities \cite{de2019depression,hosseinifard2013classifying}); and (ii) DL-based pipeline that extends CNNs \cite{seal2021deprnet,sharma2023depcap,gao2023classification,chen2021convolutional}, Transformers \cite{shao2024achieving,wang2024gctnet} and GNNs \cite{lu2024mast,li2023gcns,dai2023classification,zhu2023classification} to identify depression-related dynamics and relationships within/among different EEG channels, or brain structural/behavioural abnormalities in MRI scans.

In particular, a large part of existing expressive behaviour-based ADA approaches were developed on the AVEC depression challenge series \cite{valstar2013avec,valstar2014avec,valstar2016avec,ringeval2017avec,ringeval2019avec} organised since 2013. These challenges provided several mainstream publicly available audio-visual depression datasets facilitating researchers to share and compare their ADA solutions, including AVEC 2013 and AVEC 2014 audio-visual depression datasets containing human face videos and audio (in German language) files, as well as DAIC-WOZ dataset \cite{gratch2014distress} containing de-identified audio signals, facial primitives and language transcripts (used for AVEC 2016, AVEC 2017 and AVEC 2019 challenges). Other audio-visual datasets including CMDC \cite{zou2022semi}, PCD \cite{dibekliouglu2018dynamic}, LaB \cite{lin2021looking}, and D-vlog \cite{yoon2022d} also have been made publicly available, where the LaB dataset additionally provides human body movements. All these datasets (except D-vlog \cite{yoon2022d}) were recorded under pre-defined tasks or interviews, and provided rigorous clip-level depression labels. Besides, the EATD-Corpus \cite{shen2022automatic} recorded audio and text transcript files while the MODMA dataset \cite{cai2022multi} recorded EEG and audio signals. Due to ethical considerations, to the best of our knowledge, most MRI/EEG and expressive behaviour  datasets for ADA research are not publicly available.

Given the numerous ML-based ADA approaches developed in the past decade, this paper conducts a comprehensive survey for up-to-date ML-based automatic depression assessment (ADA) approaches building on various internal human brain behaviours recorded by EEG and MRIs, as well as external human expressive speech, facial, body and language behaviours. While early ADA surveys \cite{gao2018machine,pampouchidou2016depression,he2022deep,aleem2022machine} fail to include up-to-date advanced ADA approaches, recent surveys suffer from the following limitations: 
(i) reviewing only speech-based ADA solutions \cite{wu2023automatic}; 
(ii) reviewing audio and language-based solutions without providing comprehensive reviews for visual ADA approaches \cite{squires2023deep}; and (iii) summing ADA approaches that analyse social media cues, ignoring audio, video, MRI and EEG-based ADA solutions \cite{hasib2023depression}. 
More importantly, none of these surveys provided comprehensive psychological/physiological/medical evidence linking depression to various human internal and external behaviours. 
Our survey starts with reviewing the biological, physiological and psychological evidence linking depression with a wide variety of human internal signals and external behavioural cues (Sec. \ref{subsec:depression behaviours}), which provide a solid theoretical basis for developing human internal and external cue-based ADA approaches (surveyed in Sec. \ref{sec:existing methods}). 
In addition, publicly available ADA datasets and competitions are discussed in Sec. \ref{sec: dataset and challenge}. Finally, Sec. \ref{sec:discuss} discusses the challenges that are faced by existing ADA approaches and highlights opportunities for future research.



\section{Background}
\label{sec:background}

\noindent This section first summarises the negative impacts of depression (Sec. \ref{subsec:negative impacts}). We then review depression-related human internal and external cues/signals in Sec. \ref{subsec:depression behaviours}, which provide the inspiration and theoretical basis for researchers to develop existing ML-based ADA approaches. Fig. \ref{fig:depression_behaviour} also illustrates these impacts and behaviours.

\subsection{Health and social impacts of depression}
\label{subsec:negative impacts}

\begin{figure}
\centering
\includegraphics[width=8.8cm]{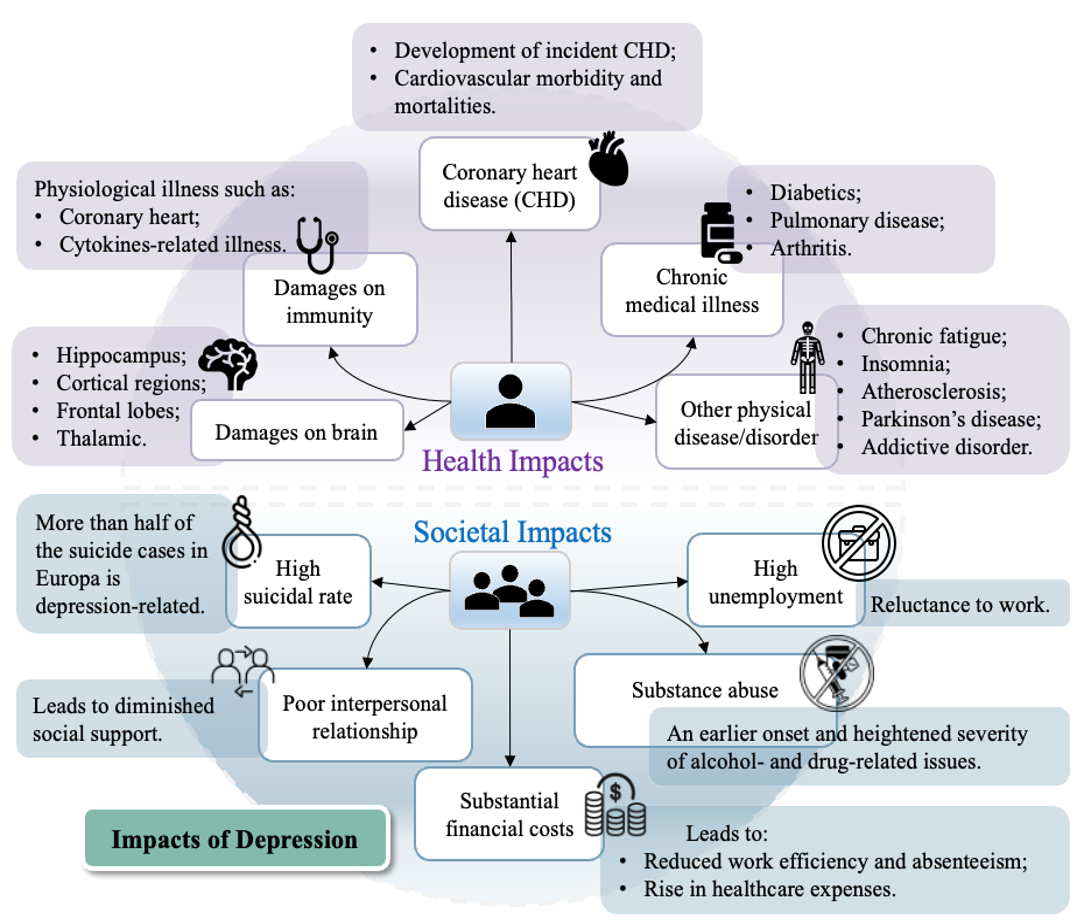}
\caption{The negative impacts of depression.}
\label{fig:negative}
\end{figure}

\noindent As summarised by Fig. \ref{fig:negative}, depression affects both human health and their social functioning. 

\textbf{(1) Health impacts:} Depression frequently causes damages on human brain regions \cite{siegle2002can}, particularly in the hippocampus, cortical regions, frontallobes, and thalamic \cite{yang2023thalamus}. It also affects human immune systems \cite{eyre2012neuroplastic}, leading to physiological illness such as coronary heart diseases \cite{gan2014depression} and cytokines-related illness \cite{yirmiya2000illness}. 
A systematic survey \cite{katon2007association} concludes that depressed chronically ill patients (e.g., diabetes, pulmonary disease and arthritis) suffer from significantly higher number of medical symptoms compared to non-depressed patients. Depression is also a high risk factor leading to peripheral vascular disease \cite{d2010depressive},  
which is also linked to chronic fatigue \cite{abbey1991chronic}, insomnia \cite{benca2008insomnia}, atherosclerosis \cite{isingrini2011early}, and Parkinson's disease \cite{hemmerle2012stress}. More seriously, depression is a potential risk factor for the initiation of cancers \cite{reiche2004stress}. The psychological intervention treating depression can prolong survival of cancer patients with different types of cancers \cite{andersen2010biobehavioral} while the meta-analysis conducted in \cite{satin2009depression} concluded that depression is a predictor of mortality in cancer patients.

\textbf{(2) Societal impacts:} Depression also leads to serious social problems. More than half of the suicide cases in Europe are linked to depression-related mental disorders \cite{hegerl2016prevention}, with another study \cite{dong2019prevalence} revealed similar findings, i.e., the combined lifetime prevalence of depression and suicide attempts is 31\%.
Li et al. \cite{li2022predictors} surveyed 24 studies including 954,882 patients with depression, suggesting that depressed patients who experienced adverse environmental factors were more prone to suicidal ideation and attempting suicide. Depression also typically leads to poor interpersonal relationships \cite{stoetzer2009problematic,kupferberg2023social}. The inadequate interpersonal relationships, in turn, signify diminished social support and contribute to suicide-related behaviours \cite{lewinsohn1997depression}. Besides, 
addictive disorders are also frequently presented within depressed adolescents \cite{rao2006links}.
Moreover, depressed individuals frequently suffer from substantial financial costs in the form of time away from work and a rise in healthcare expenses \cite{thapar2022depression}, who are more susceptible to reduced workplace efficiency \cite{lepine2011increasing}
diminished earnings/job loss, homeless and increased substance abuse and poverty \cite{kupferberg2023social}.


\subsection{Depression-related human behaviours}
\label{subsec:depression behaviours}

\begin{figure*}
\centering
\includegraphics[width=16.9cm]{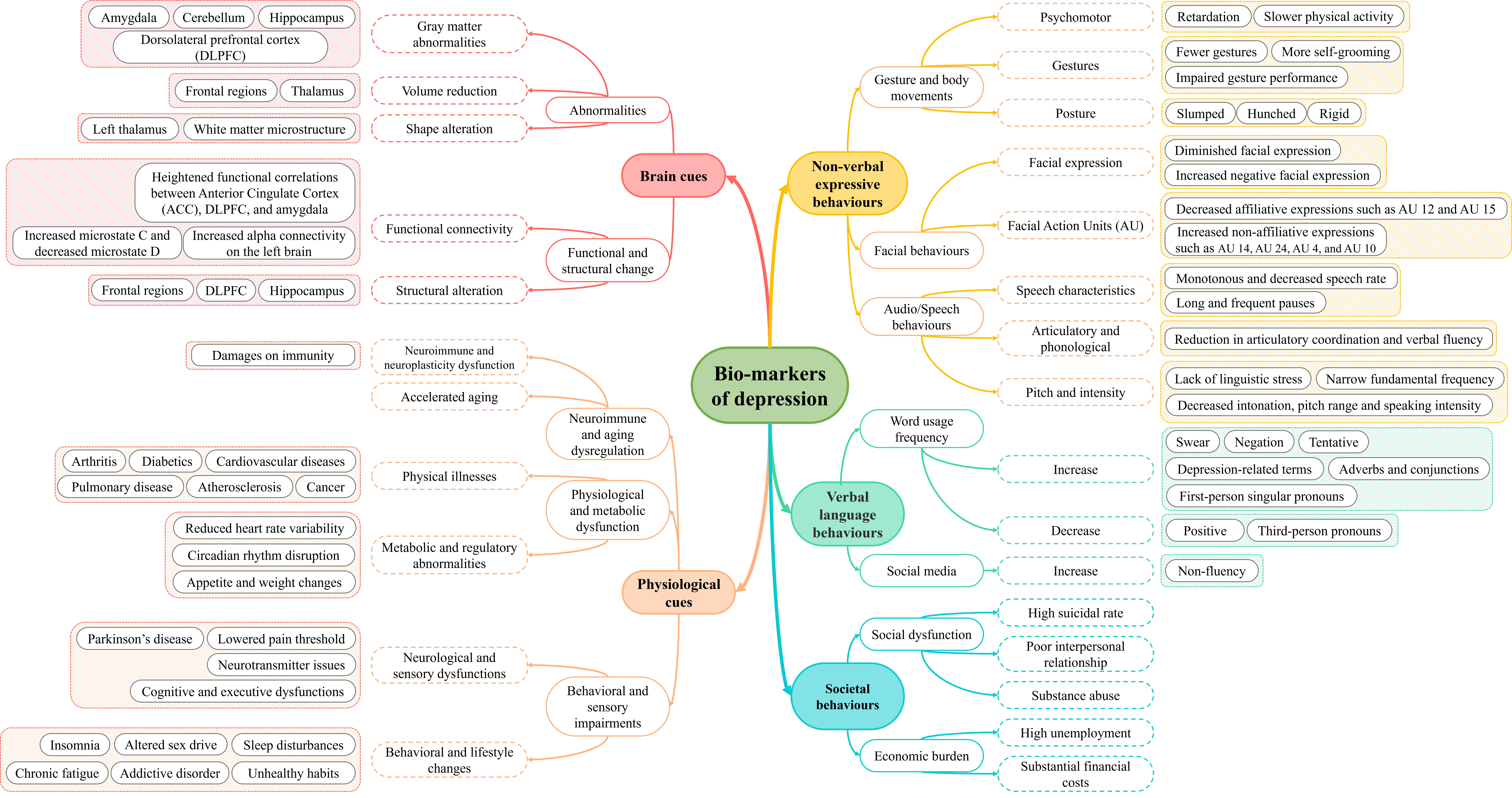}
\caption{Illustration of common human behavioural biomarkers associated with depression.}
\label{fig:depression_behaviour}
\end{figure*}

\noindent Previous studies found that depressed and non-depressed individuals frequently show different internal and external behaviours (illustrated in Fig. \ref{fig:depression_behaviour}), which form the theoretical basis for developing existing ADA systems.


\subsubsection{Brain cues}
\label{subsec:brain_behaviour}


\noindent Depressed individuals typically experience abnormal brain processes. This is evidenced by brain regions implicated in the pathophysiology of depression \cite{mayberg2000regional}. In particular, magnetic resonance imaging (MRI) have been widely employed to explore brain activities related to depression, revealing that depression would seriously impair several brain regions, including abnormal gray matter changes of dorsolateral prefrontal cortex (DLPFC) \cite{siegle2002can}, amygdala \cite{lu2012impaired} and cerebellum \cite{he2017co}, as well as damage symptoms of hippocampus and thalamus (significantly smaller volumes \cite{van2013paralimbic,yang2023thalamus}).
Some EEG studies found that depressed individuals usually exhibit hemispheric asymmetry in their brain signals compared to non-depressed individuals \cite{gotlib1998eeg}, as they usually have more alpha activity at the left side of their brains \cite{lee2018neurophysiological}. 
Besides,  Li et al. \cite{li2021altered} apply microstate analysis on EEG data of depressed and healthy individuals, suggesting that depression leads to increase in microstate C and decrease in microstate D, which further convinces depression's negative influences on brain alpha band.

\subsubsection{Physiological cues}

\noindent Depression is also associated with abnormal physiological states, as changes in the brain affects on various body systems. 
%
For instance, fluctuations in serotonin that are related to depression frequently cause depressed individuals suffering from sleep problems \cite{benca2008insomnia} and lower sex drive \cite{staley2006sex}.
%
The state of `accelerated aging' \cite{wolkowitz2010depression} is another typical depression biomarker, as depression causes comparable changes on human brain's plasticity and synaptic functions  \cite{sibille2013molecular} that are similar to those occurring with aging, where depressed individuals usually suffer from 4–6 years of accelerated aging compared to normal people \cite{verhoeven2014major}. 
%

\subsubsection{Physical indicators}

Meanwhile, Katon et al. \cite{katon2003clinical} and Lin et al. \cite{lin2011effects} concluded that depression leads individuals to experience decreased heart rate variability, increased adhesiveness of platelets, negative health habits such as smoking and over-eating, and low circadian rhythmicity in their skin body temperatures \cite{barbini1998perceived}. 
Appetite change is another physical indicator of depression, where 
greater appetite change tends to indicate greater depression severity \cite{paykel1977depression}. 
As a result, human body shape and weight changes are important physical indicators to indicate depression \cite{polivy1976clinical}, e.g., the obesity \cite{milaneschi2019depression}, where a reciprocal and reinforcing cycle of maladaptive physiological adjustments can establish an interconnection between the depression and obesity \cite{fu2023shared}. 
%
%
As discussed in Sec. \ref{subsec:negative impacts}, sickness-like behaviours, cognitive dysfunction, and anhedonia \cite{eyre2013treating} are also typical depression-related cues. For example, there is a bidirectional relationships linking depression to the progression of cancer \cite{giese2011decrease} and linking the persistent inflammatory processes to stresses in the manifestation of depressive symptoms \cite{sotelo2014biology}.



\subsubsection{Non-verbal expressive behaviours} 
\label{subsec:expressive behaviours}

Depressed individuals typically express different non-verbal expressive behaviours, which facilitates existing audio-visual ADA approaches. 

\textbf{Gesture and body movements:} Depressed individuals typically show physical activity known as psychomotor retardation \cite{gupta2009major}, represented by different gait from normal people \cite{deligianni2019emotions} such as slumped, hunched, or rigid posture that restricts body movements \cite{france2001communication}), impaired gesture movements (e.g., gesture deficits and degraded intransitive behaviours) \cite{pavlidou2021hand}, fewer gestures 
during interpersonal interactions \cite{hinchliffe1975study}, as well as downwards head-tilting and frequent 
self-adaptor, self-touch or hand-over-face gestures \cite{mahmoud2016towards,mahmoud2011interpreting,lin2021looking}.


\textbf{Facial behaviours:} A typical facial depression biomarker is the reduced or even diminished facial expressions \cite{gupta2009major,gaebel2004facial} and head movements \cite{joshi2013can}, especially significantly reduced expressions of positive emotions.
However, contradicting conclusions have been reached in terms of negative facial expressions, where some studies claimed that depressed individuals experience increased negative expressions \cite{ellgring2007non} while others shown that even negative facial expressions are less expressed by them \cite{gaebel2004facial}.
From a more fine-grained perspective, Cohn et al. \cite{cohn2009detecting} and Girard et al. \cite{girard2014nonverbal} conducted pioneering investigations regarding the depression-related facial AU patterns, both of which show that participants suffering from high depression severity display less affiliative facial expressions (AU 12 and AU 15) but more non-affiliative facial expressions (AU 14). Also, an early study \cite{waxer1974nonverbal} suggested that depressed individuals tend to avoid eye contact, and show more of nonspecific gaze and less mutual gaze, indicating lack of interest in communication.

\textbf{Audio behaviours:} Non-verbal audio behaviours are also crucial depression biomarkers \cite{tolkmitt1986effect}, as depression-related cognitive impairments would affect human working memory that controls the articulatory system and speech-based information storage \cite{murphy1999emotional}. Specifically, depressed individuals typically exhibit monotonous speech \cite{gupta2009major} characterised by frequent pauses with a `muffled' voice \cite{silva2021voice}, reduction in articulatory coordination \cite{williamson2013vocal} and verbal fluency \cite{husain2020cortical}, abnormal pitch variability, vocalization time, fundamental frequency and formants \cite{mundt2012vocal,nilsonne1988measuring}, long pause times \cite{mundt2012vocal}, as well as an overall decrease in speech rate/speed \cite{hollien1980vocal}.  An early study summarised depressive voices as `patients speak in a low voice, slowly, hesitatingly, monotonously, sometimes stuttering, whispering, try several times before they bring out a word, become mute in the middle of a sentence' \cite{kraepelin1921manic}.
Besides, Hollien et al. \cite{hollien1980vocal} found that depressed individuals are more likely to experience reduced speaking intensity, pitch range and intonation, as well as a lack of linguistic stress, while Breznitz et al. \cite{breznitz1992verbal} revealing that narrower frequencies are expressed by depressed females' speeches compared to health females. 

\subsubsection{Verbal language behaviours}
\label{subsec:verbal language behaviours}

\noindent Language-based communication is a primary mean for people to exchange their ideas and intentions, and thus its styles are strongly associated with their depression status \cite{nilsonne1988measuring}, despite that there exists cross-cultural differences in depression-related verbal expressions \cite{Tsugawa2015RecognizingDF} (e.g., specific negative emotions or phrases are stronger depression indicators in Chinese social platforms than in other western platforms \cite{wang2013depression}). Among them, the most recognised verbal depression biomarkers are the increase usage of the first person singular (i.e., `I', `me', and `mine') \cite{weintraub1989verbal}, swear words and depression terms, as well as the decrease usage of 3rd person pronoun (e.g., `She', `He'). These are evidenced by the studies conducted on suicidal poets \cite{stirman2001word}, student essays \cite{rude2004language}, Twitter \cite{de2013predicting} and a Chinese social platform - Sina \cite{wang2013depression}. A survey \cite{jh2012verbal} further revealed that individuals with depression and suicidal tendencies have an increased usage of self-referential words such as 'I', 'me', and 'mine'. They additionally found that depressed individuals have a general decrease  usage of 3rd person pronoun (e.g., 'She', 'He'), and a increase over time for swear word use and frequency of depression terms. 
Other studies \cite{de2013social} further found that tentative words (e.g., maybe, perhaps), negations (e.g., no, never), adverbs, and conjunctions are more commonly used by depressive individuals, with less usage of verbs, while postpartum depression mothers tending to show more non-fluency in their social platform posts. From another perspective, Rude et al. \cite{rude2004language} found that depressed individuals used more negatively valenced words, while Cummins et al. \cite{cummins2015review} summarised that they are more likely to express dull, misery and “lifeless” descriptors. 
Besides, the sentences expressed by depressed patients are usually fragmented or shortened, reflecting their reduced coherence and cognitive difficulty \cite{Resnik2015BeyondLE}.


\subsubsection{Other depression-related human indicators}

In addition to the behaviours discussed above, depressed patients also show various cognitive deficits compared to non-depressed ones, including varied exploratory behaviours \cite{blanco2013influence}, impaired recognition of basic emotional facial expressions \cite{bourke2010processing}, longer reaction time, a decreased psychomotor ability \cite{marazziti2010cognitive} and diminished capability in acquiring new information (but retaining the capability in gathering information) \cite{marazziti2010cognitive}.
Memory abnormality is a key sign reflecting depression, as depression causes the suppression of hippocampal neurogenesis, inhibition of dopamine neurons, and increased sensitivity of the amygdala \cite{dillon2018mechanisms}. As evidenced by \cite{ramponi2004recollection}, depressed individuals tend to exhibit difficulties in recalling information, i.e., they frequently experience enhanced memory for negative content but poorer memory for positive material \cite{burt1995depression}. 
With such cognitive and memory deficits and bias, multiple human cognitive capabilities are further negatively impacted, i.e., depressed individuals frequently show abnormality or lower performances in decision-making \cite{blanco2013influence}, problem-solving \cite{elderkin2006executive}, planning \cite{elliott1997prefrontal} behaviours, as well as spending time with similarly depressed others and pair-wise interactions rather than group interactions \cite{elmer2020depressive}.



\section{Survey Method}

\noindent This section presents our strategy to identify the papers included in the survey by reporting the procedure, search query, inclusion/exclusion criteria, selection process and publication analysis, while a structured four-phase flow figure for reporting items of our survey guided by \cite{moher2010preferred} is illustrated in Fig. \ref{fig:PRIMA}. Example search queries and the terminologies used in this survey are provided in the Supplementary Material.

\begin{figure}
\centering
\includegraphics[width=8.8cm]{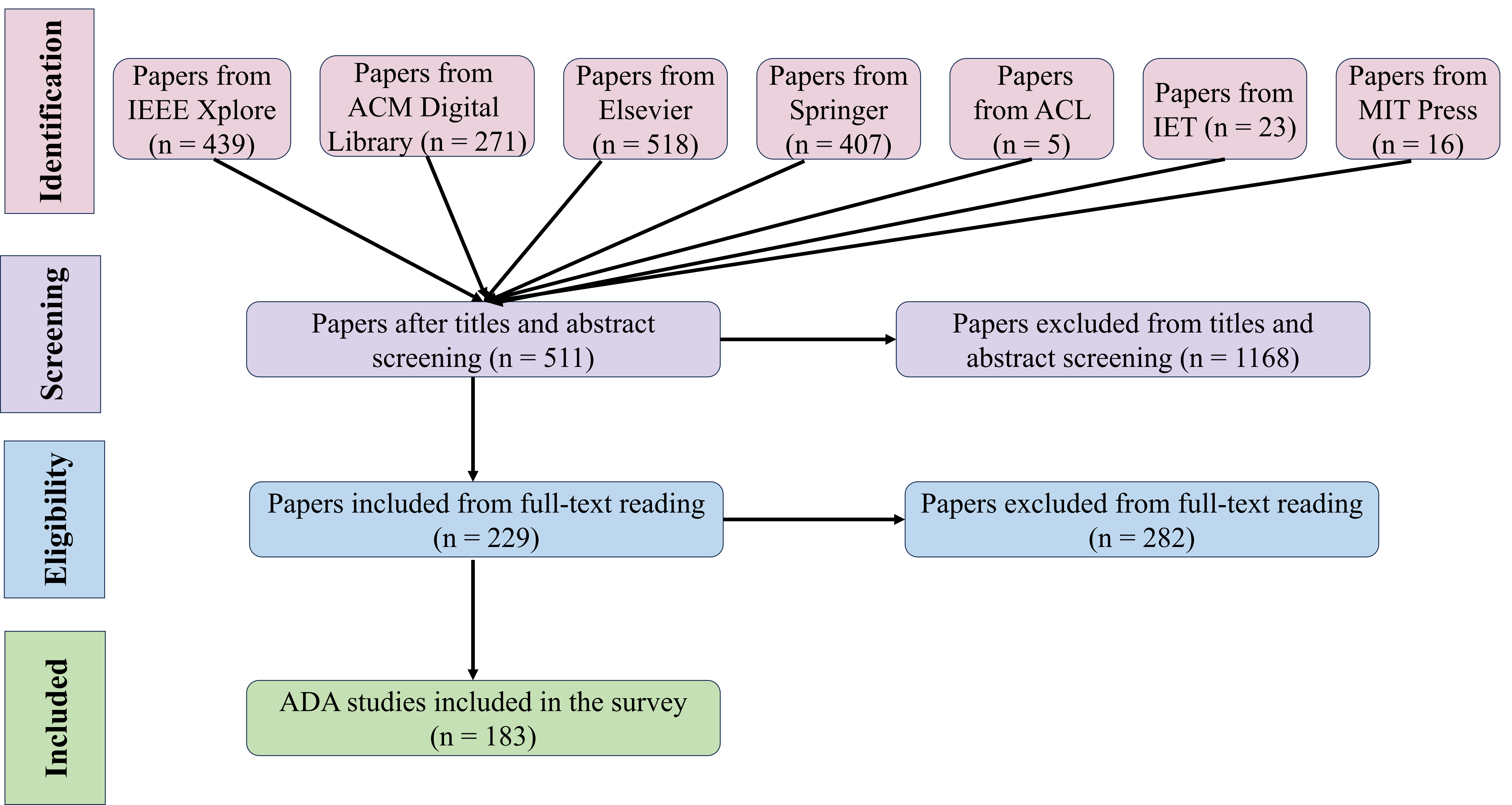}
\caption{PRIMA Schema for our systematic review.}
\label{fig:PRIMA}
\end{figure}



\textbf{Procedure:} we follow \cite{nightingale2009guide} and \cite{moher2010preferred} to systematically review previous ADA publications, ensuring the quality and replicability of this survey, as well as the assessment of each study’s strengths and weaknesses.

\textbf{Search Query:} We gathered publications based on the terms found in their titles, abstracts and keywords, utilising the IEEE Explore, ACM Digital Library, Elsevier, Springer, ACL Anthology, IET and MIT Press, as their broad coverage in affective computing, medical data analysis and computer science. These search queries were adjusted slightly to accommodate the requirements of each database. 

\textbf{Eligibility Criteria:} While it is impossible to exhaustively summarise all ADA approaches in this survey, \textit{papers were included only if:} (i) publication dates range between 2012 and 2024; (2) they provide adequate description of ADA algorithms, with clear experimental settings and results; (3) their title, abstract or keywords contain at least one keyword describing ADA-related technology and one keyword from the Search Query Keywords; (4) they propose new ADA solutions rather than only applying them; and (5) they published in English. Meanwhile, \textit{papers were excluded if:} (1) they did not involve any ML algorithm for inferring depression; and (2) they were not under peer-review process or come from sources that may not meet credibility standards.

\textbf{Selection Process:} All candidate publications were initially searched based on their titles and abstracts, followed by a full-text review to ensure compliance with eligibility criteria. To establish consensus, we assessed a random sample of three papers for each modality, with full agreement on details of inclusion and exclusion criteria. The remaining papers were randomly assigned to at least two authors for independent screening, where ambiguous cases were resolved through discussion. This way, a total of 1679 papers were retrieved, with 183 were ultimately included (the full list is provided in the Supplementary Material).

\textbf{Publication analysis:} We defined a set of variables to understand key information from each ADA paper using an analytical approach, including numerical variables such as depression classification accuracy, the numbers of participants and clips, as well as categorical variables such as modality, methodology type, predictor and fusion strategy (summarised in Table \ref{tb:methods}, with the full list of variables provided in the Supplementary Material). A descriptive statistics description is employed to represent numerical variables, while categorical variables were derived through either a top-down or bottom-up strategy.

\section{Automatic Depression Assessment}
\label{sec:existing methods}


\begin{table*}[!t]
\caption{Summary of the surveyed ADA approaches, where `C' and `R' denote the classification and regression, respectively; `Hybrid' denotes approaches utilising both traditional ML and DL techniques; \textbf{IAIF:} Iterative Adaptive Inverse Filtering; \textbf{OQNN:} Open Quotient Neural Network; \textbf{DAE:} Denoising Auto-Encoder; \textbf{EOH:} Edge Orientation Histogram; \textbf{ADTP:} Audio Delta Ternary Patterns; \textbf{ReHo:} Regional Homogeneity; \textbf{FALFF:} Fractional Amplitude of Low Frequency Fluctuations.}


\tiny
\centering
\renewcommand{\arraystretch}{1.25}
\scalebox{1.1}{
\begin{tabular}{m{0.9cm}<\raggedright m{0.8cm}<\raggedright  m{0.8cm}<\raggedright  m{3.9cm}<\raggedright  m{3cm}<\raggedright  m{4cm}<\raggedright }

\hline
\centering Modality & \centering ADA task & \centering Framework & Feature extractor & Predictor & References \\ \hline
\multirow{4}{*}{EEG} & \multirow{4}{*}{C} & ML & GA, LDA, EMD, CBEM, GSW & SVM, KNN, RF, NB, DT &\cite{shen2023depression}, \cite{mohammadi2015data},  \cite{zhu2020improved},  \cite{shen2020optimal}\\
\cline{3-6}
&& \multirow{1}{*}{DL} & CNN, CNN+LSTM/GRU, MLP, GNN, Transformer & MLP &   \cite{dang2020multilayer}, \cite{wan2020hybrideegnet},  \cite{sharma2023depcap}, \cite{shao2024achieving}, \cite{wang2024gctnet}, \cite{lu2024mast}, \cite{shen2023depression}, \cite{ay2019automated}, \cite{song2022lsdd}, \cite{tigga2022efficacy}, \cite{tian2024board}, \cite{ying2024functional} \\

\cline{3-6}
&& Hybrid & GNN & SVM, MLP & \cite{zhang2024novelEEG}, \cite{li2023gcns}, \cite{wang2021identification} \\

\hline
\multirow{5}{*}{MRI} & \multirow{4}{*}{C} & ML & ReHo, FALFF, ICA, RFE, LASSO & SVM, LDA, GPC, KNN, DT, LR, RF, GNB, BNB, DT, TVF, SVF &\cite{mousavian2021depression}, \cite{gao2018machine}, \cite{winter2024systematic} \\
\cline{3-6}
&& DL & CNN, GNN & MLP &\cite{chen2021convolutional}, \cite{gao2023classification}, \cite{dai2023classification}, \cite{zhu2023classification}, \cite{yao2020temporal}, \cite{pominova2018voxelwise}, \cite{lin2023automatic}, \cite{jun2020identifying}, \cite{ktena2018metric}, \cite{fang2023unsupervised}, \cite{qin2022using}, \cite{jaiswal2019automatic}, \cite{li2024automated} \\

\cline{3-6}
&& Hybrid & CNN, LSTM & BernoulliNB, MLP & \cite{mousavian2020depression} \\
\cline{2-6}
& C \& R & DL & GNN & MLP & \cite{kong2021spatio}\\


\hline
\multirow{10}{*}{Visual} & \multirow{4}{*}{C} & ML & PCA + LDA, AAM, LBPH, LBP-TOP, HOG, RBM, GMM & KNN, SVM, GMM, RF, NBC, TF, SG, PSD & \cite{gahalawat2023explainable}, \cite{girard2013social}, \cite{anis2018detecting}, \cite{joshi2013can}, \cite{girard2014nonverbal}, \cite{harati2019classifying},  \cite{yang2023trial},  \cite{al2018depression}, \cite{yang2022clustering},  \cite{alghowinem2015cross}, \cite{wang2020gait}, \cite{wang2021detecting}, \cite{joshi2012neural}, \cite{joshi2013relative}, \cite{lu2022postgraduate} \\

\cline{3-6}
&& DL & MLP, CNN, Transformer & MLP & \cite{li2024automatic}, \cite{li2024sftnet}, \cite{xie2021interpreting}, \cite{guo2022automatic}, \cite{niu2024depressionmlp} \\

\cline{3-6}
&& Hybrid & LBP-TOP, MRLBP-TOP, CNN, CNN-LSTM & DT, SVM, KNN, DT, MLP & \cite{he2018automatic}, \cite{shao2021multi}, \cite{yu2022depression}, \cite{liu2024multimodal} \\

\cline{2-6}
& \multirow{4}{*}{R} & ML & LPQ, LGBP-TOP, MRLBP-TOP, LPQ-TOP & SVR & \cite{tao2019low}, \cite{du2019encoding}, \cite{dhall2015temporally}, \cite{wen2015automated}, \cite{song2024loss} \\
\cline{3-6}



&& \multirow{1}{*}{DL} & MLP, CNN, CNN+RNN, GRU, CNN+Transformer, GNN &  
MLP &\cite{de2021mdn}, \cite{xu2024two}, \cite{pan2023integrating}, \cite{ahmad2021cnn}, \cite{shang2021lqgdnet}, \cite{song2022spectral}, \cite{he2022depnet}, \cite{niu2024pointtransform}, \cite{jaiswal2019automatic}, \cite{liu2022measuring}, \cite{zhao2023novel},  \cite{li2020depression}, \cite{zhou2018visually}, \cite{uddin2020depression}, \cite{pan2024spatial}, \cite{al2018video}, \cite{chen2021sequential}, \cite{zhu2017automated}, \cite{zhang2023mtdan}, \cite{he2024depressformer}, \cite{de2019combining}, \cite{moreno2023expresso}, \cite{chen2022neural}, \cite{pan2024opticaldr}, \cite{wang2024facialpulse}, \cite{li2024automatic} \\


\cline{2-6}
& \multirow{2}{*}{\makecell[l]{C \& R}} & DL & Transformer & MLP & \cite{zhang2023improved} \\
\cline{3-6}
&& \multirow{1}{*}{Hybrid} & CNN & LightGBM, SVR, MLP & \cite{song2018human}, \cite{islam2024facepsy} \\

\hline
\multirow{9}{*}{Audio} & \multirow{3}{*}{C} & ML & MFCC, CLP & QDA, LR, GMM, MLC & \cite{mundt2012vocal}, \cite{moore2007critical}, \cite{ozdas2004investigation} \\
\cline{3-6}
&& DL & CNN, CNN+LSTM & MLP & \cite{vazquez2020automatic}, \cite{ma2016depaudionet} \\

\cline{3-6}
&& Hybrid & COVAREP, CNN, LSTM & RF, DT, MLP & \cite{zhang2021depa}, \cite{wu2024mobile} \\
\cline{2-6}
& \multirow{4}{*}{R} & ML & & RF, LR & \cite{sun2017random} \\
\cline{3-6}
&& DL & CNN, LSTM, Transformer & MLP & \cite{he2018automated}, \cite{al2018detecting}, \cite{amiriparian2017snore}, \cite{zhao2019automatic} \\

\cline{3-6}
&& Hybrid & MFCC, ADTP, rPPG, LGBP-TOP, HRV, CNN, CNN-LSTM & RF, MLP & \cite{casado2023depression}, \cite{fu2022audio} \\

\cline{2-6}
& \multirow{2}{*}{\makecell[l]{C \& R}} & ML & & SVM, RF & \cite{valstar2016avec} \\
\cline{3-6}
&& DL & CNN, RNN/LSTM & MLP & \cite{othmani2021towards}, \cite{rejaibi2022mfcc}, \cite{muzammel2020audvowelconsnet}, \cite{zhang2021depa} \\

\hline
\multirow{3}{*}{\makecell[l]{Langua-\\ ge/text}} & \multirow{2}{*}{C} & ML & N-gram, LDA, LIWC & LR, SVM, RF, AdaBoost, DT, UDL, TF-IDF & \cite{de2013predicting}, \cite{de2013social}, \cite{nguyen2014affective}, \cite{Tsugawa2015RecognizingDF}, \cite{shen2017depression}, \cite{Tadesse2019DetectionOD}\\
\cline{3-6}
&& DL & CNN, GRU, Transformer & MLP & \cite{trotzek2018utilizing}, \cite{perez2024longitudinal}, \cite{khowaja2024depression} \\
\cline{2-6}
& R & ML & Unigrams, LIWC, LDA & LR, SVR &  \cite{Resnik2015BeyondLE}, \cite{resnik2013using} \\
\hline
\multirow{12}{*}{\makecell[l]{Multi-\\ modal}} & \multirow{4}{*}{C} & ML & HMMs, AMMs, FACS & KNN, SVM, DT, NB & \cite{pampouchidou2016depression}, \cite{cohn2009detecting}, \cite{pampouchidou2015designing}, \cite{nasir2016multimodal}, \cite{senoussaoui2014model}, \cite{alghowinem2016multimodal}, \cite{ning2024depression} \\
\cline{3-6}
&& DL & CNN, LSTM/GRU, Transformer & MLP & \cite{yoon2022d}, \cite{shen2022automatic}, \cite{ye2021multi}, \cite{cheong2024fairrefuse}, \cite{wei2022multi}, \cite{zheng2023two}, \cite{shao2024multimodal}, \cite{tao2024depmstat}, \cite{ye2024dep}, \cite{jung2024hique} \\

\cline{3-6}
&& Hybrid & Mfcc, DAE, COAVREP, FACS & SVM, RF, LR, MLP & \cite{lin2021looking}, \cite{dibekliouglu2018dynamic}, \cite{small_but_fair} \\
\cline{2-6}
& \multirow{4}{*}{R} & ML & LBP, EOH, LPQ, MHH, COVAREP, LIWC & RF, PLS, SVR, RF &\cite{gong2017topic}, \cite{valstar2013avec}, \cite{valstar2014avec}, \cite{ringeval2017avec}, \cite{gratch2014distress}, \cite{jan2014automatic}, \cite{gupta2014multimodal}, \cite{meng2013depression}, \cite{kachele2014inferring} \\
\cline{3-6}
&& DL & CNN, LSTM/GRU, CNN+LSTM, Transformer, GNN & MLP &\cite{he2021automatic}, \cite{niu2020multimodal}, \cite{rodrigues2019multimodal}, \cite{ringeval2019avec}, \cite{he2022reducing}, \cite{kaya2019predicting}, \cite{hong2022using}, \cite{uddin2022deep}, \cite{hu2024parallel}, \cite{sun2022cubemlp}, \cite{sun2022tensorformer}, \cite{yin2019multi}, \cite{fang2023multimodal}, \cite{zhang2024mddr}, \cite{fan2024transformer}, \cite{gimeno2024reading}
\\

\cline{3-6}
&& Hybrid & CNN, GNN & SVR, PLS, MLP & \cite{jan2017artificial}, \cite{shen2024multi} \\

\cline{2-6}
& \multirow{4}{*}{\makecell[l]{C \& R}} & ML & FAUs, dMFCCs, COVAREP, IAIF, OQNN & SVM, DT & \cite{yang2016decision}, \cite{williamson2016detecting}, \cite{scherer2013audiovisual} \\
\cline{3-6}
&& DL & CNN+LSTM, GNN, Transformer  & MLP & \cite{muzammel2021end}, \cite{zhang2024novel}, \cite{chen2023semi}, \cite{Lam2019ContextawareDL}, \cite{zhang2024multimodal} \\

\cline{3-6}
&& Hybrid & CNN & SVM, RF, MLP & \cite{yang2017hybrid}, \cite{yang2018integrating}, \cite{haque2018measuring} \\
\hline
\end{tabular}}
\label{tb:methods}
\end{table*}

\noindent This section comprehensively surveys existing ADA approaches that rely on human brain behaviours (i.e., via EEG and various MRI data), as well as expressive facial, body, audio and language behaviours.


\subsection{EEG-based automatic depression assessment}

\noindent Non-invasive electroencephalography (EEG) is a popular sensor for monitoring human brain electrical activities that are related to human depression status. Thus, EEG-based ADA solutions have been extensively developed.

\subsubsection{Challenges and solutions}

While a significant depression-related EEG biomarker is its spectral features, whose principal bands can be divided into five parts: Delta, Theta, Alpha, Beta, and Gamma describing different physiological meanings,  \textbf{a key challenge unique to EEG-based systems is determining which bands to use for ADA}. Hosseinifard et al. \cite{hosseinifard2013classifying} found that the combination of alpha and theta bands is a discriminative depression feature. When logistic regression (LR) is applied to process such features, 90\% depression classification accuracy is achieved for 90 test subjects. 
\textbf{Other challenges unique to EEG-based ADA, including: (i) how to properly model the spatial information derived from multiple EEG channels; and (ii) the varying importance of different channels \cite{shen2023depression,zhang2024novelEEG}.} In this sense, Shen et al. \cite{shen2020optimal} modify kernel target alignment (KTA) strategy to optimally choose a subset of important EEG channels (addressing problem (ii)) for reducing the information redundancy, which are then processed by a CNN for EEG-based ADA. Their later work \cite{shen2023depression} develops an adaptive channel fusion framework to model the spatial information derived from multi-channel EEG signals (addressing problem (i)), where two improved focal loss functions are employed to upweight the loss values for the hard examples. In addition, recent studies \cite{zhang2024novelEEG,li2023gcns,wang2021identification} frequently treat EEG signals as graphs to address these three problems, where each EEG channel is represented as a node, with edges are defined to capture relationships between connected EEG channels. 
Subsequently, GNN with an attention module 
\cite{zhang2024novelEEG}, or Multi-scale Adaptive Spatial-Temporal GNN \cite{lu2024mast} 
have been proposed to weight and jointly process spatial cues of multiple EEG channels. We further discuss other EEG-based ADA methods in the following subsection.

%


\subsubsection{Other EEG-based ADA systems}
To extract depression-related features from EEG signals, Mohammadi et al. \cite{mohammadi2015data} utilises Linear Discriminant Analysis (LDA) and Genetic Algorithm for the feature extraction. These features are then fed to a decision tree (DT) to predict the depression score. 
While hand-crafted features for EEG-based ADA are tedious, time consuming \cite{ay2019automated} and require domain knowledge, recent studies \cite{zhu2020improved,dang2020multilayer,wan2020hybrideegnet,tian2024board,ying2024functional,kwok2025machine} also deep learn depression-specific features from EEG signals. 
For example, Seal et al. \cite{seal2021deprnet} propose a CNN to process the pictorial representation of the given EEG signal for ADA. Betul et al. \cite{ay2019automated} combine CNN and Long-short-term-memory (LSTM) network to process EEG signals obtained from left and right hemispheres of human brain for assessing depression level. 
Similar CNN-LSTM architecture also has been proposed by \cite{song2022lsdd} for EEG-based ADA. 
Aiming at real-time EEG-based ADA, Sharma et al. \cite{sharma2023depcap} first compute a spectrogram image from the target EEG signal using STFT, and then feed it to a CNN-LSTM or CNN-Gated Recurrent Units (GRU) network for classification. 
A couple of transformer networks have been extended for EEG-based ADA, including attention-based gated recurrent units time-series transformer model \cite{tigga2022efficacy}, Decentralised-Centralised Structure Transformer \cite{shao2024achieving}, and Graph Convolutional Transformer \cite{wang2024gctnet}. 
These approaches directly applied existing transformer architecture for EEG-based ADA but 
did not investigate how to design depression-specific modules.


\subsection{MRI-based depression assessment}

Brain images obtained from functional MRI (fMRI), structural MRI (sMRI), resting-state fMRI (rsfMRI) and diffusion MRI (diffusion tensor images, DTI), manifest crucial aspects of human brain functions related to depression (detailed discussed in Sec. \ref{subsec:brain_behaviour}), and thus also are informative data source for inferring depression status.

\subsubsection{Challenges and solutions}

\textbf{The common challenges in MRI-based ADA includes how to model spatio-temporal nature of the MRI data, and identifying the conductivity patterns/regions relevant to depression within MRI data.} To model spatio-temporal cues, CNN models \cite{mousavian2020depression,chen2021convolutional,li2024automated} have been introduced to learn depression-specific spatio-temporal cues from MRI images, e.g., Pominova et al. \cite{pominova2018voxelwise} employ multiple deep voxelwise networks to assess depression on structural (3D) and functional (4D) MRI data, each of which is built on several residual volumetric convolutional modules. Alternatively, an unsupervised discrepancy-based cross-domain fMRI adaptation framework is proposed \cite{fang2023unsupervised}, which applies attention-guided GCN to model spatio-temporal fMRI patterns for ADA. Similarly, Yao et al. \cite{yao2020temporal} first segment each region-of-interest (ROI)-based fMRI image series and then propose a Temporal-Adaptive GCN to model depression-related spatio-temporal cues. 
%
For the latter challenge, Gao et al. \cite{gao2023classification} first define a heatmap of conductivity patterns indicating positive influences within the brain, which then guides a lightweight 3D CNN in learning depression cues from 3D Gray Matter data recorded by fMRI, where patterns reflected by prefrontal cortex, insula, superior temporal cortex, and cingulate cortex have been identified to be linked with depression. Qin et al. \cite{qin2022using} employ GCN to process rsfMRI for depression assessment, showing that the most salient regions related to depression are within the default mode, fronto-parietal, and cingulo-opercular networks, while Nodal topologies of the left inferior parietal lobule is also associated with depression symptoms.
%
We discuss other MRI-based ADA methods in the subsection below.

\subsubsection{Other MRI-based ADA systems}

Gao et al. \cite{gao2018machine} review 63 approaches and conclude that a common pipeline for MRI-based ADA approaches typically consists of three stages: (i) MRI data pre-processing, (ii) depression feature extraction, and (iii) ADA model training/inference. A comparative study \cite{winter2024systematic} systematically evaluates the feasibility of applying traditional ML models to analyse biomarkers provided by sMRI, fMRI and DTI images for ADA, where various traditional ML models (e.g., SVM, KNN, RF, LR, naive Bayers and boosting classifier) were employed, with diagnostic classification accuracy ranged between 48.1\% and 62.0\%. Recently, spatial GCN \cite{dai2023classification,jun2020identifying,zhu2023classification} and spatio-temporal GCN \cite{kong2021spatio} have been found to be particularly effective for MRI-based ADA. 
Ktena et al. \cite{ktena2018metric} propose a Siamese GCN that adopts spectral graph convolutions to process fMRI data for ADA. It encodes each fMRI time series as a graph with topology decided by Pearson Correlation Coefficient (PCC) computed between time-series channels, where a novel graph similarity metric is introduced to measure the similarity between each pair of fMRI graphs during training.
%
%
%
Meanwhile, Mousavian et al. \cite{mousavian2021depression} propose a unified framework for sMRI and rsfMRI-based ADA, including data preparation (e.g., removing backgrounds and data normalisation), 
3D MRI region definition, whole-brain matter similarity computation based on pre-defined 3D regions, 
spectral domain feature extraction, Wilcoxon rank-sum test-based feature selection, and naive Bayers-based depression classification. 
%

\subsection{Face/head-based depression assessment}
\label{subsec:facial behaviour ADA}

Since human facial/head behaviours provide informative depression biomarkers (discussed in Sec. \ref{subsec:expressive behaviours}) while recent advanced DL models can accurately and automatically recognise various facial/head behaviours (e.g., AUs, head movements and facial expressions), automatic visual behaviour-based depression assessment is becoming a prevalent research topic in the past decade. Existing facial/head behaviour-based ADA approaches can be categorised into three types: (i) face image-based approaches which assess depression merely based on a static facial display (e.g., a video frame) \cite{valstar2013avec,valstar2014avec}; (ii) video-based approaches which assess depression based on spatio-temporal facial behaviours \cite{xu2024two,de2021mdn,harati2019classifying,tao2019low,he2022depnet,du2019encoding}; and (iii) facial attribute-based approaches which assess depression based on pre-extracted anonymous facial descriptors (e.g., facial landmarks \cite{pan2023integrating,niu2024pointtransform}, AUs \cite{song2018human,song2022spectral,jaiswal2019automatic}, head poses \cite{valstar2016avec,ringeval2017avec}, gazes/blinks \cite{yang2022clustering,alghowinem2015cross} or frame-level deep facial features \cite{ringeval2019avec}).

\subsubsection{Challenges and solutions} 

\textbf{(i) a key challenge for face-based ADA is that depression is better reflected by spatio-temporal facial behaviours rather than static images}, i.e., depressed and non-depressed individuals can express same facial displays in a single frame. However, some ADA approaches still model frame-level depression-related facial cues despite the availability of face videos, which frequently aggregate frame-level predictions via average or majority voting for video-level assessment. For example, the baselines and participants of AVEC 2013 \cite{valstar2013avec} and AVEC 2014 challenges \cite{valstar2014avec}, generally extract different hand-crafted image-level descriptors (e.g., Histogram of oriented gradients (HOG) \cite{dalal2005histograms} and Local Phase Quantisation (LPQ) \cite{ojansivu2008blur}), and then feed them to SVM Regression (SVR) for assessing frame-level depression severity. Following a similar pipeline, several early studies also extract Gabor wavelet features \cite{girard2013social,jan2014automatic} from static face images, with KNN as the depression predictor. However, \textit{such approaches re-use the video-level label as the frame-level labels to train models, and thus may cause `ill-posed problems' (i.e., similar static facial expression inputs paired with different depression labels), making it impossible to learn good hypothesis.} To address this issue, spatio-temporal modelling strategies have been particularly explored for ADA (please refer to Sec. \ref{subsec:facial_video_method} for details); 
\textbf{(ii)} while depression is associated with long-term and multi-scale facial behaviours \cite{song2022spectral}, \textbf{most existing spatio-temporal ADA methods only learn depression cues from short segments, ignoring crucial long-term (i.e., video-level) but variable-length spatio-temporal depression cues at the feature extraction level \cite{al2018video,pan2024spatial}.} This is partially addressed by video-level modelling ADA approaches. Song et al. \cite{song2022spectral,song2018human} propose a spectral encoding strategy which treats the facial primitives as a multi-channel time-series, applying Fourier Transform to generate a raw spectral representation, and then normalises its size through frequency alignment to standardise all frame-level features as a fixed-size video-level representation. Building on this, Chen et al. \cite{chen2022neural} extend the method by incorporating a Neural Architecture Search (NAS) in order to identify optimal networks for inferring depression status from the video-level facial primitive spectral representations. Another study \cite{chen2021sequential} investigates a sequential chained-fusion strategy to extract complementary depression patterns contained in both facial display and facial dynamics for the video-level depression assessment. Melo et al. \cite{de2021mdn} 
propose a maximization branch to learn multi-scale smooth facial variations and a difference branch for encoding multi-scale facial dynamics from the video-level image sequence, both of which are fused for video-level ADA; \textbf{(iii) the third limitation is that most existing ADA methods were devoted to assessing adults' depression status.} To bridge this gap, Abbasi et al. \cite{abbasi2022statistical} construct a video-level spatio-temporal graph  
for assessing children's depression status; 
\textbf{(iv) another challenge for inferring depression from facial/head behaviours is that these behavuours are more prone to data privacy and ethical issues.}
%
%
Thus, some ADA studies \cite{song2022spectral,niu2024pointtransform,valstar2016avec,valstar2013avec} were developed to infer depression from anonymous facial attributes (discussed in Sec. \ref{subsec:facial_primitive_method}). Alternatively, Pan et al. \cite{pan2024opticaldr} design a privacy preservation ADA system which records de-identified facial images, while retaining not only depression-relevant features but also emotion-related cues for face image-level ADA; and \textbf{(v) face/head data can also suffer from noisy and ambiguous depression annotations.} He et al. \cite{he2022reducing} proposes three modules (self-attention, square ranking regularization and re-label modules) to reduce the impact of noisy and ambiguous depression annotations. To deal with missing faces, face detection errors, frame loss and incorrect labels, Song et al.  \cite{song2024loss} introduces a generic loss function relaxation strategy to mitigate abnormal loss values for various loss functions during video-based ADA model training. We discuss 
other existing face/head behaviour-based ADA approaches in the following sub-paragraphs.

\subsubsection{Face image-based approaches}

while depressed and non-depressed individuals can express same static facial displays (e.g., neutral face), some approaches still attempt to assess depression from a single face image (e.g., facial expression and pupil \cite{liu2024multimodal,li2024automatic}). 
Melo et al. \cite{de2019depression} propose a distribution learning strategy to train a ResNet for face image-based ADA. 
Liu et al. \cite{liu2022measuring} pre-train a CNN network on a facial expression dataset and then linearly superimpose the entropy of the detected expression into depression severity.
Zhao et al. \cite{zhao2023novel} propose a frequency attention strategy to identify depression from static facial displays.
%
Li et al. \cite{li2024automatic} integrate a dual-scale convolution and a channel-wise adaptive attention to a standard CNN network, aiming to infer depression from a facial expression. 
%
Zhang et al. \cite{zhang2023improved} propose a two-stream deep network LKCT. 
A decoupled knowledge distillation module then distills the LKCT into the lightweight network for face image-based ADA. 
%
%
%
%
%
Similarly, a standard frame-level DL-based ADA pipeline also learns depression cues from individual face frames and aggregates predictions or features for video-level assessment.
Zhou et al. \cite{zhou2018visually} propose a DepressNet consisting of two convolution layers and four bottleneck residual convolution blocks to identify depression-informative facial regions from each face frame for ADA. Shang et al. \cite{shang2021lqgdnet} propose to extract each facial image with local textures in the quaternion domain to preserve the dependence of color channels, while deep learning global facial features via a simple CNN. Then, it fuses hand-crafted local features with the deep-learned global cues for frame-level depression assessment. To model multi-scale spatial cues, He et al. \cite{he2021automatic} propose local and global attention blocks to capture depression-related facial features at multiple spatial scales, which are then combined via a Weighted Spatial Pyramid Pooling for image-level ADA. 
%
%



\subsubsection{Face video-based approaches} 
\label{subsec:facial_video_method}
Facial temporal dynamics contain crucial biomarkers (discussed in Sec. \ref{subsec:expressive behaviours}) for inferring depression. 
Early face video-based ADA approaches frequently summarise video-level spatio-temporal facial behaviours by computing either the video-level statistics of hand-crafted frame/segment-level facial features (e.g., the major facial expressions, variance of facial expressions \cite{scherer2013automatic}) or differences/dynamics between facial frames such as velocity of angular \cite{girard2014nonverbal}, facial appearance eigenviectors \cite{cohn2009detecting}, three orthogonal planes-based dynamic representations (e.g., LBP-TOP, LGBP-TOP, LPQ-TOP) \cite{joshi2013can,gupta2014multimodal,kaya2014ensemble,he2018automatic,dhall2015temporally}, Motion History Histogram (MHH) \cite{meng2013depression}, and pixel differences between adjacent frames \cite{kachele2014inferring}. 
For instance, Pampouchidou et al. \cite{pampouchidou2015designing} extract three types of features, including appearance-based features (LBP and LBP-TOP), model-based features (active appearance models (AAMs)), and motion-based features (i.e., frame differencing, feature point tracking and optical flow), from every frame/short segment of the given face video, and then feed their statistics to a SVR for ADA. 
He et al. \cite{he2018automatic} extend the LBP-TOP feature to MRLBP-TOP for extracting short-term dynamics and then apply Fisher Vector to aggregate them for video-level ADA. SVR then integrates decisions across dimensions for video-level ADA.
%
%
Instead of directly learning depression cues from face videos, Casado et al.\cite{casado2023depression} calculate over 60 video-level statistical, geometrical, and physiological features from the remote photoplethysmography signals obtained from each face video for ADA.
Yang et al. \cite{yang2023trial} proposes a Trial Selection Tensor Canonical Correlation Analysis to fuse spatio-temporal relevance between facial dynamic features and pupil diameter features for ADA due to their complementary affective nature.


A prototypical DL solution for face video-based ADA is through learning frame-level facial features/predictions and aggregating them as a video-level depression representation.
Uddin et al. \cite{uddin2020depression} apply LSTM to model the facial dynamics from CNN-learned frame-level depression features. 
Xu et al. \cite{xu2024two} first encode a multi-scale facial spatial representation from every frame of the target video, and then integrate all frame-level multi-scale spatial features as a single spectral graph representation for video-level ADA in an effort to learn a richer and more condensed latent representation. 
Pan et al. \cite{pan2024spatial} integrate both spatial and temporal attention mechanisms to 3D CNNs, 
and then combine all segment-level depression predictions as the video-level prediction. 
Similarly, Al et al. \cite{al2018video} utilise a C3D network to extract short segment-level facial dynamics and then averages all segment-level predictions for video-level ADA. 
Li et al. \cite{li2024sftnet} and Melo et al. \cite{de2019combining} propose two-stream networks that jointly learns facial image-level depression cues and video-level facial dynamics, both of which are concatenated via a  Multi-Layer Perceptron (MLP) for the final video-level ADA. 
Zhu et al. \cite{zhu2017automated} propose two-stream networks that jointly learns facial appearance features and short-term facial dynamics via optical flow. 
%
Zhang et al. \cite{zhang2023mtdan} propose a lightweight CNN incorporating temporal difference for short-term facial behaviours, followed by an attention for long-term facial behaviour modelling, which are fused via Interactive Multi-head Attention for video-level ADA. He et al. \cite{he2024depressformer} employ a video Swin-Transformer to learn spatio-temporal and fine-grained depression-specific features, integrating them via a Depression Channel Attention Fusion (DCAF) module for video-level ADA. Xie et al. \cite{xie2021interpreting} design an end-to-end ADA framework for variable-length videos, which leverages a 3D CNN for local temporal pattern extraction and a Redundancy-Aware Self-Attention (RAS) scheme for global feature aggregation.
%
%
%
%
Moreno et al. \cite{moreno2023expresso} pre-train a CNN on action recognition datasets and fine-tune it on ADA face videos, enabling analysis of frame-level regions and short-term temporal expressions.

\subsubsection{Facial behaviour primitive-based approaches} 
\label{subsec:facial_primitive_method}
Due to ethical concerns, original face images/videos sometimes are not available, as faces reveal identity while depression status is sensitive. 
Thus, some ADA studies are developed to infer depression from anonymous facial attributes, where frame-level facial landmarks, discrete facial expressions, Action Units (AUs), head poses/motions, and eye gaze directions and blinks are the most prevalent primitives. 
%
%
For instance, Pan et al. \cite{pan2023integrating} apply CNN to learn a video-level representation from facial landmarks, and then propose a spatio-temporal attention network to extract its depression-related features for ADA. 
%
%
Niu et al. \cite{niu2024pointtransform} first divide the video-level facial landmark sequence into several segments along the temporal dimension, and then extend the MLP-mixer \cite{tolstikhin2021mlp} to make a depression prediction from every segment. 
%
%
%
%
Kacem et al. \cite{kacem2018detecting} utilise GMM model and Fisher vector encoding to represent Lie-algebra based rotation matrix of 3D head motions. 
Based on eye blinks, Al et al. \cite{al2018depression} summarise video-level eye blink-related features such as blink rate and duration from facial landmarks, and then individually employ SVM, Naive Bayes and Adaboost for binary depression classification. 
%
Since a systematic study investigating the relationship between AUs and depression \cite{cohn2009detecting} suggests that AUs can provide vital information for depression assessment, a certain number of ADA systems \cite{song2022spectral,song2018human,gong2017topic} also attempt to extract informative depression cues from AUs, e.g.,
Muzammel et al. \cite{muzammel2021end} utilise a MLP to process 20 AUs, while Gong et al. \cite{gong2017topic} computing statistics of AUs appeared in every interview topic for for video-level ADA.

Instead of using only a single primitive, more approaches utilise multiple facial primitives for ADA. The baselines of AVEC 2016 and AVEC 2017 \cite{valstar2016avec,ringeval2017avec} feed various frame-level facial primitives (e.g., AUs, head poses, gaze directions and discrete emotions) to linear SVM or RF for frame-level depression classification orseverity estimation. Then, the video-level depression classification or severity estimation is achieved via temporal majority voting, averaging frame-level predictions, or aggregating facial primitive statistics. Song et al. \cite{song2018human} also compute 12 statistics 
to summarise all frame-level AUs/head poses/gaze directions of the given video, which are concatenated and processed by SVR for video-level ADA. 
Guo et al. \cite{guo2022automatic} propose a temporal dilated CNN to extract long-term video-level facial dynamics from frame-level facial landmark and head pose sequences, and then apply an attention operation to adaptively weight different feature map channels for video-level ADA. 
Niu et al. \cite{niu2024depressionmlp} propose MLPs with a Dual Gating to process facial landmarks and AUs of the given video for depression severity estimation, where the Dual Gating maintains the physical properties of input landmarks and AUs while a Mutual Guidance module interacting facial landmarks and AUs in the context of their global information. 
Recently, Islam et al. \cite{islam2024facepsy} develop an mobile  
system for real-world ADA, where segment-level statistics of facial landmarks, AUs, gazes and head poses are fused and selected for inferring depression. 

\subsection{Body behaviour-based depression assessment}

\noindent Human body behaviours, especially gait behaviours, are distinguishable between depressed and non-depressed individuals \cite{wang2020automatic,wang2021detecting}. Some studies \cite{wang2020gait,wang2021detecting} identified that \textbf{silhouette extraction is a key challenge unique to body behaviour-based ADA}. To address this, they learn depression-related time-domain features for analysing gait abnormalities, frequency-domain features describing video-level body dynamics and frame-level spatial geometric features, which are finally concatenated and fed to a MLP or SVM for ADA. 
%
%
Besides, Shao et al. \cite{shao2021multi} leverage LSTM to process sequential skeleton behaviours while two CNNs are trained to extract silhouette features from the gait video. The predictions of both modalities are combined to make the final depression prediction. 
%
%
Liu et al. \cite{liu2022measuring} develop a two-stream 3D CNN to predict depression severity from human body movements during walking, i.e., the motion amplitude computed from the 18 joints' positions of the target individual. These positions are fused with deep-learned spatio-temporal facial expression cues for video-level ADA. 

In addition to gait, limb movements are also important biomarker for assessing depression. 
Joshi et al. \cite{joshi2012neural} compute spatio-temporal cues from human upper body key points
via a temporal visual words dictionary which are fused with LBP-TOP feature-based facial behaviour codebook learned from the video, where MLP and SVM-based classifiers are employed to make the final depression assessment. 
Their later work \cite{joshi2013relative} further explores relative orientation and radius of the human upper body in addition to the upper body key points, which are then fused and processed by SVM for ADA. 
Yu et al. \cite{yu2022depression} propose a spatial attention dilated TCN (SATCN) which applies a hierarchy group of temporal convolution operations with various dilated convolution scales to learn human skeletal cues for ADA. 
Lu et al. \cite{lu2022postgraduate} extract a rigid-body representation by analysing kinetic energy and potential energy collected from the human walking video, and then apply fast Fourier transform to encode them as a joint energy feature, where SVM, RF and KNN are individually employed as the classifier. 
Ahmad et al. \cite{ahmad2021cnn} individually apply various 2D CNNs to assess depression from either human upper body frames or facial frames, and then averaging all frame-level predictions to achieve video-level ADA, which suggests that face and upper body frames provide similar ADA performance. 
Lin et al. \cite{lin2021looking} leverage various body behaviours for ADA by extracting frame-level body pose features using OpenPose. 
From these, they derive multiple body behaviour features, including video-level statistics, self-adaptor features reflecting individual body manipulation, and fidgeting behaviour features indicating unconscious or semi-conscious movements. A GMM and Fisher Vector encoding are then used to summarise frame-level features into a video-level representation for ADA.


\subsection{Audio-based depression assessment}
\label{subsec:audio behaviour ADA}

\noindent While human non-verbal audio behaviours contain abundant depression-indicative information \cite{cummins2015review}, three types of audio features have been widely explored for ADA, including: (i) \textit{acoustic spectral and frequency features} (e.g., spectral, frequency, voice quality and glottal features); (ii) \textit{prosodic features} measuring quasi-periodic rate of vocal fold vibration, speaking rate, pausing and intonation patterns \cite{moore2007critical}; and (iii) \textit{deep-learned features} capturing human uninterpretable but depression-specific audio cues. 

\subsubsection{Challenges and solutions}

To extract hand-crafted audio features for ADA, \textbf{(i) feature pre-processing is challenging as these features may possess high dimensionality.} Jan et al. \cite{jan2017artificial} apply PCA to the extracted MFCC features along with various spectral LLDs (e.g., spectral flatness, spectral flux, and other spectral features). Zhang et al. \cite{zhang2024novel} vertically concatenate visual representations of complementary MFCCs and mel spectrogram extracted from original audio waveform to form a matrix, which is subsequently processed by a 2D-CNN, an attention mechanism, and a BiLSTM for ADA. Alternatively, DL-based approaches (discussed in Sec. \ref{subsec:deep_audio_system}) have been proposed to tackle the \textbf{(ii) unaddressed problem posed by hand-crafted features, i.e., they can not capture subtle and human uninterpretable depression indicators within audio signals}. Besides, \textbf{(iii) aggregating frame/segment-level audio feature as a clip-level audio depression representation is another key research question \cite{shen2017depression}}, where statistics \cite{moore2007critical}, GMM \cite{williamson2016detecting}, temporal convolution \cite{ye2021multi}, temporal attention \cite{pan2023integrating}, LSTM \cite{zhang2021depa,ma2016depaudionet,al2018detecting}, majority voting \cite{vazquez2020automatic}, and GNNs \cite{chen2023semi} have been extended to address this challenge.
%
%
We detailed survey the existing audio-based ADA methods within the subsections below.


\subsubsection{Hand-crafted audio feature-based approaches} 

Mel-frequency cepstral coefficients (MFCCs) \cite{zhang2024novel,othmani2021towards,he2018automated,rejaibi2022mfcc,jan2017artificial,gupta2014multimodal,fu2022audio,perez2024longitudinal} is one of the most widely adopted acoustic spectral and frequency features for ADA, which are derived from spectrograms by capturing the short-term energy spectrum and revealing subtle changes in energy overtime, such as monotony, slower speech rate, and increase pauses (related to depression) \cite{kraepelin1921manic}. 
As a result, a large number of studies \cite{chen2023semi,yang2017hybrid,williamson2016detecting} leveraged MFCCs along with other hand-crafted acoustic low-level descriptors (LLDs) for ADA. 
%
%
The results validate that the MFCCs are the best individual indicator for ADA among different audio spectral and frequency features. 
%
Speeches expressed by depressed individuals are typically characterised by lacking in expression and reduced variability in fundamental frequency and prosody \cite{kraepelin1921manic}, voice perceptual qualities, prosodic/vocal tract features (e.g., movement of articulators), pitch (emotional and physiological changes in speeches \cite{hollien1980vocal,gupta2009major}) as well as glottal waveform  relating to the volume–velocity of airflow through the glottis from the lungs \cite{ozdas2004investigation}.
%
Consequently, these acoustic features have been specifically employed by audio/multi-modal ADA systems \cite{zhang2024novel,muzammel2020audvowelconsnet,sun2017random}. 
Moore et al. \cite{moore2007critical} extract prosodic, vocal tract, and glottal source features at the frame level to analyse audio recordings from English-speaking participants reading a short story for at least three minutes. Clip-level representations are then obtained using mean, standard deviation, and temporal summaries of these frame-level features. 
A quadratic discriminant analysis (QDA)-based classifier finally performs depression classification using each audio clip-level representation. Results indicate that the combination of prosodic and glottal features yields the highest accuracy. 
Williamson et al. \cite{williamson2016detecting} utilise a formant tracking algorithm \cite{mehta2012kalman} to estimate F1, F2 and F3 from each normalised audio clip in every 10 ms. Meanwhile, they employ peak-to-root-mean-square (RMS) to measure loudness, and openSMILE to extract MFCCs on a segment-level basis. All these features are jointly fed to a Gaussian staircase regressor to learn the entire audio clip-level depression representation. 
%
%
%
Similarly, Alhanai et al. \cite{al2018detecting} also apply COVAREP to extract frame-level formant, pitch, glottal and voice quality features, which are fed to a three-layer LSTM for audio clip-level ADA.

\subsubsection{Deep learning approaches}
\label{subsec:deep_audio_system}

DL-based approaches are frequently achieved through three strategies: 
(i) leveraging/fine-tuning existing pre-trained DL networks; 
(ii) developing specific DL networks for preliminary depression-related audio feature extraction; 
and (iii) developing specific DL networks to further refine extracted hand-crafted audio features. 
In particular, the DeepSpectrum \cite{amiriparian2017snore,zhao2019automatic}, SincNet \cite{ravanelli2018speaker}, and NetVLAD \cite{arandjelovic2016netvlad} are the most widely-used pre-trained DL networks for audio-based ADA, as they provide strong emotion-related audio representations. An typical example is proposed by \cite{ye2021multi}, which applies DeepSpectrum to convert raw audio waveform as a set of segment-level spectrograms, and then extract segment-level deep features via a pre-trained VGG-16. The clip-level depression representation is finally aggregated via a Temporal Convolutional Network (TCN) with an attention layer. 
Pan et al. \cite{pan2023integrating} adopt the pre-trained SincNet to reduce the amount of model training needed to extract frame-level audio features.
Then, a Temporal Attention (TA) block adaptively weights such frame-level features along the temporal dimension, which are further fed to an unaligned transformer to generate the clip-level audio representation for ADA. 
%
Instead of using pre-trained models, Zhang et al. \cite{zhang2021depa} 
trained a self-supervised DEPA model using the Switchboard dataset \cite{godfrey1992switchboard}. 
The trained DEPA processes pre-extracted audio Mel-scale spectrograms through CNN layers, followed by an LSTM for clip-level ADA. 
A similar approach \cite{vazquez2020automatic} also utilises segment-level audio spectrograms as the input to a 1D-CNN, where the majority voting is applied to aggregate segment-level predictions for clip-level depression classification to address the clip-level aggregation challenge.
%
%
Othmani et al. \cite{othmani2021towards} propose an EmoAudioNet which starts with extracting frame-level MFCCs and segment-level spectrogram features, and then individually processes them via a MFCC-based CNN and a spectrogram-based CNN, whose outputs are finally concatenated for depression assessment.
Chen et al. \cite{chen2023semi} split each audio signal into a set of one-second segments, from each of which MFCCs and other hand-crafted LLD features are extracted. These segment-level audio features are subsequently processed by a GNN-based Semi-Supervised Domain Adaptation model (GNN-SDA) for ADA. In \cite{ma2016depaudionet}, each extracted frame-level Mel-scale filter bank feature first undergoes a standard normal variate normalization, and are then concatenated along time-axis as a set of segment-level 2D time-frequency representations, which are further processed by a CNN-LSTM model and a SGD regressor to make segment-level depression predictions.

\subsubsection{Hybrid approaches.} 

To leverage advantages of both hand-crafted and deep-learned features for a more effective audio-based ADA, 
Kaya et al. \cite{kaya2019predicting} utilise the extended Geneva minimalistic acoustic parameter set (eGeMAPs) \cite{eyben2015geneva} to extract frame-level LLDs (e.g., breathing, fillers, lip noise, laughter) and apply DeepSpectrum to learn segment-level audio features. These feature sets are then combined for ADA, with results highlighting the complementarity of hand-crafted and deep-learned audio features in ADA. He et al. \cite{he2018automated} extract 38 hand-crafted LLDs capturing spectral, cepstral, prosodic and voice quality cues, while deep learning (via a CNN) features from spectrograms and waveforms of the given audio clip. Based on them, semantic median robust extended local binary patterns (MRELBP) are further computed. All features are concatenated and passed to a FC layer for ADA.
Yang et al. \cite{yang2017hybrid} extract hand-crafted frame-level MFCCs, loudness, energy, pitch, jitter and shimmer features from the given audio, and further summarise them as a clip-level statistical representations. Finally, a CNN is employed to learn the clip-level depression score. Rejaibi et al. \cite{rejaibi2022mfcc} extract MFCCs, 53 hand-crafted acoustic features (e.g., fundamental frequency (F0), formants, jitter and shimmer). These frame-level features are then fed to an RNN model to learn their temporal dependencies and produce the clip-level depression prediction.


\subsection{Language/text-based depression assessment}
\label{subsec:language behaviour ADA}

\noindent Language/textual data (e.g., conversations, social media posts, and clinical notes) provide semantic meaningful verbal information that can directly and individually identify human depression status \cite{coppersmith2015adhd}.


\subsubsection{Challenges and solutions}

As discussed in Sec. \ref{subsec:verbal language behaviours}, \textbf{the key challenge for language/text-based ADA is that some verbal depression biomarkers are language-dependent.} Consequently, Chen et al. \cite{chen2023semi} developed separate models (i.e., Chinese BERT and English BERT models) to generate sentence-level embeddings from texts of different domains. \textbf{Another common challenge is utilising multi-scale (word-level, sentence-level and topic level) contexts when analysing depression-related language cues,} where a frequently employed solution is to extract and combine multi-scale languages features \cite{de2013predicting,de2013social,Tadesse2019DetectionOD,shen2017depression}, or extract word/sentence-level features by specifically considering conversation/text topics \cite{Lam2019ContextawareDL,trotzek2018utilizing,gong2017topic}. We detailed review language/text-based ADA approaches in the subsection below.


%


\subsubsection{Hand-crafted approaches} 

Hand-crafted language features for ADA can be categorised into three primary scales: \textbf{lexical} (word-level), \textbf{syntactic} (fragmented sentence-level), and \textbf{semantic} (topic-level). Each scale provides unique insights into language use and psychological depression indicators. Specifically, \textbf{lexical features} are extracted from individual words to describe their linguistic or emotional attributes (e.g., pronouns, verbs, negative emotion words, and anger words), where Pennebaker’s Linguistic Inquiry and Word Count (LIWC) \cite{pennebaker2001linguistic} is a prevalent feature extraction strategy for text-based ADA \cite{resnik2013using,de2013predicting,shen2017depression,coppersmith2015adhd,nguyen2014affective,Tadesse2019DetectionOD,de2014characterizing,gratch2014distress}. Additionally, more tailored lexicons such as word count, pronoun usage, and other stylistic markers (detailed discussed in Sec. \ref{subsec:verbal language behaviours}) that are critical for depression assessment also have been widely utilised \cite{de2013predicting}. Meanwhile, \textbf{syntactic features} focus on the structural patterns of language at the fragmented sentence-level, as fragmented sentences expressed by depressed individuals often reflect reduced coherence and cognitive difficulty \cite{Resnik2015BeyondLE}. These include the number of replies, shares, or interactions indicating individuals' engagement in social media contexts \cite{de2013predicting}, i.e., reduced social interaction or lower mention rate have been proven to be strong depression indicators \cite{de2013predicting,Tsugawa2015RecognizingDF}. Besides, \textbf{semantic features} operate at a more broader level to capture semantics and themes provided by languages. A widely-used depression semantic feature is extracted through Latent Dirichlet Allocation (LDA) \cite{blei2003latent}, which identifies latent topics reflecting patterns of language use and thematic areas associated with depression (e.g., expressions of loneliness, hopelessness and mental health concerns \cite{resnik2013using,shen2017depression,Tsugawa2015RecognizingDF,nguyen2014affective,Tadesse2019DetectionOD,Resnik2015BeyondLE}). 


As these three types of linguistic features capture complementary depression-related information at different scales, they have been frequently combined for ADA, with a large part of them being applied to social media data. 
For example, Choudhury et al. \cite{de2013predicting} analyse Twitter data from the year preceding depression onset by extracting lexical features (e.g., LIWC, as well as terms related to anxiety, insomnia and antidepressant mentions), syntactic features (e.g., reply proportion and question-focused posts), and semantic features (e.g., post frequency, shares, posting time and social graph metrics) characterizing social interactions. Finally, an SVM integrates these features for ADA. 
Its extension \cite{de2013social} further monitors population-level depression trends in real-time, additionally analysing unigrams (an n-gram consisting of a single item from a sequence, such as a single word) and bigrams (a pair of consecutive written units such as letters, syllables, or words) from Twitter posts to account for the general language use, and then applies a PCA and SVM for more accurate ADA. 
%
%
Shen et al. \cite{shen2017depression} construct a publicly available Twitter dataset, from which three types of features including LIWC emotional linguistic features, semantic LDA topic features, and social interaction features such as the numbers of tweets, followers and followings, are extracted, which are then jointly encoded into a sparse representation and processed by a regularised LR model for ADA. 
Based on LIWC features and user engagement metrics, Coppersmith et al.  \cite{coppersmith2015adhd} incorporate open-vocabulary methods to specifically character n-gram language models, for capturing depression condition-specific character-level linguistic features. Besides Twitter, texts appeared in other social media channels have also been utilised for ADA. 
Tadesse et al. \cite{Tadesse2019DetectionOD} develop their approach on a Reddit posts dataset annotated with binary depression labels. It adopts LIWC, LDA, and n-gram modeling to extract features related to correlation significance, hidden topics and word frequencies. The three textual feature sets are then concatenated and fed to various ML classifiers for ADA. 
Apart from social media contents, 
Resnik et al. \cite{resnik2013using} apply LIWC and LDA to extract lexical and semantic features from student essays. It categorises words into psychological groups and employs topic modeling to enhance depression feature extraction. Their subsequent work \cite{Resnik2015BeyondLE} extends LDA as a supervised nested LDA (SNLDA) to specifically extract depression-related semantic contents for ADA. Based on LiveJournal, Nguyen et al. \cite{nguyen2014affective} extracts LIWC features, LDA features, affective features, and mood tags selected by users for their posts, where a regularised regression model is employed to refine features and make binary depression decisions. 
Sun et al. \cite{sun2017random} analyse DAIC-WOZ interview transcripts to define six depression-related attributes (e.g., sleep quality, PTSD diagnosis, treatment history, personal preferences, emotions). They then identify five most depression-related attributes, and fuse their predictions using a cascade strategy with a RF model to produce the final depression severity.
Tsugawa et al. \cite{Tsugawa2015RecognizingDF} employ bag-of-words to extract word-level linguistic features (e.g., emotional tone, word frequency), while LDA derives semantic content features. Additionally, social interaction metrics (retweet rate, mention rate and URL usage) are further extracted. These features are concatenated into a single vector and processed via SVM for ADA. 
%

\subsubsection{Deep learning approaches} 
A typical DL-based language ADA solution  \cite{shen2022automatic} adopts ELMo to extract sentence-level embeddings and then feed them to a DL network consisting of two BiLSTM layers and an attention layer. This results in a contextualised text representation that emphasises the most depression-relevant sentences, which is then concatenated with audio features via a FC layer for depression assessment. 
%
%
%
Hong et al. \cite{hong2022using} propose a schema-based GNN which schematises underlying associations among words (i.e.,  node embeddings), aiming to capture depression symptoms within the entire transcript. Alternatively, DL-based approaches also frequently model semantics related to question topics and responses of the given language/textual data. Lam et al. \cite{Lam2019ContextawareDL} identify every topic from transcripts and generate augmented data by shuffling topic segments within selected combinations. The augmented texts are then used to train a transformer for deep depression-related textual feature extraction. During inference, the extracted features are integrated with features of other modalities via a dense classification layer for ADA. Based on a Chinese dataset recorded individuals' conversation with physician, Ye et al. \cite{ye2021multi} employ Word2Vec for initial word embedding extraction, and a one-hot Transformer for deep feature extraction. These features are finally concatenated with features from audio modality for ADA. 
Based on posts and comments from Reddit, Trotzek et al. \cite{trotzek2018utilizing} propose two approaches. 
The first applies LIWC to extract hand-crafted features capturing linguistic and psychological characteristics, which are then fed into a LR classifier. The second applies fastText and GloVe to learn word-level embeddings to model semantic relationships in the text, followed by a CNN to learn depression-specific cues. 
%
%
Alhanai et al. \cite{al2018detecting} apply Doc2Vec \cite{doc2vec} to generate an embedding of every textual responses, which further understands the varying prognostic relevance to depression from them. These features are finally processed by LSTMs and FC layers for ADA.

\subsection{Multi-modal automatic depression assessment}

\noindent Since different human behaviour modalities could provide complementary cues for depression assessment \cite{liu2024multimodal,small_but_fair,zheng2023two}, a large number of approaches attempted to leverage multi-modal human behaviours for ADA.

\subsubsection{Challenges and solutions} 
\textbf{A common challenge unique to multi-modal ADA systems is when to fuse multi-modal features.}  Typical solutions include combining frame/segment-level multi-modal features \cite{uddin2022deep,yoon2022d,zhang2024mddr}, fusion of multi-modal features at multiple temporal scales \cite{hu2024parallel,shen2024multi}, fusion of clip-level predictions achieved for all modalities \cite{senoussaoui2014model,valstar2016avec,ringeval2017avec,williamson2016detecting,pampouchidou2016depression}, and fusion of multi-modal clip-level representations \cite{dibekliouglu2018dynamic,scherer2013audiovisual,alghowinem2016multimodal,tao2024depmstat,niu2020multimodal,pampouchidou2016depression,gupta2014multimodal,nasir2016multimodal,zhang2024novel,wei2022multi,zhang2024multimodal}. 
\textbf{How to adaptively combine multi-modal features/predictions is also a key challenge,} where average of multi-modal predictions \cite{valstar2013avec,valstar2014avec,senoussaoui2014model,valstar2016avec,ringeval2017avec}, multi-modal feature concatenation \cite{dibekliouglu2018dynamic,scherer2013audiovisual,pampouchidou2016depression,gupta2014multimodal,nasir2016multimodal,yang2018integrating,rodrigues2019multimodal}, Gaussian Staircase Model \cite{williamson2016detecting}, cross-modal attentions \cite{niu2020multimodal,zhang2024novel,zhang2024multimodal}, graph \cite{shen2024multi} and other DL operations \cite{uddin2022deep,cheong2024fairrefuse,hu2024parallel,wei2022multi,lin2021looking} have been frequently utilised.



%
%


%



\subsubsection{Audio-visual approaches} 

\textbf{Hand-crafted approaches:} Early multi-modal ADA solutions also frequently rely on different combinations of hand-crafted features and traditional ML predictors. The baselines of AVEC 2013 and AVEC 2014 challenges \cite{valstar2013avec,valstar2014avec} extract various hand-crafted image-level static and spatio-temporal facial descriptors (discussed in Sec. \ref{subsec:facial behaviour ADA}), as well as apply openSMILE \cite{eyben2010opensmile} to extract frame-level LLD audio features. Then, clip-level audio and visual depression predictions are obtained by separately feeding clip-level audio and visual representations (i.e., mean vectors across all frame-level features) to SVR models, followed by a decision-level fusion. 
Senoussaoui et al. \cite{senoussaoui2014model} extract i-vector representations from frame-level MFCC features as the clip-level audio features while averaging all LGBP-TOP features summarising short-term facial behaviours as the clip-level visual representation, based on which unimodal audio/facial depression occurrence or severity are individually predicted via traditional ML predictors (e.g., SVM), where the decision-level fusion is conducted to make the final clip-level multi-modal prediction. 
Dibekliouglu et al. \cite{dibekliouglu2018dynamic} first apply Stacked Denoising Autoencoders (SDAE) to encode all frame-level facial expression and head motions that are related to depression status, and then summarise them as a fixed-size clip-level representation via IFV coding and Compact Dynamic Feature Set. Meanwhile, clip-level audio features including fundamental frequency, switching pause duration, Maxima Dispersion Quotient and Peak Slope are also extracted. These clip-level features are finally enhanced by a feature selection, and combined to be processed by LR for ADA. 
Scherer et al. \cite{scherer2013audiovisual} extract three clip-level audio features, including differentiated glottal flow, quasi-open quotient and dyadic wavelet transform feature. These features are concatenated with two clip-level facial features: emotional variability computed from all frame-level discrete facial expressions and motor variability describing hand gesturing and head movements, which are finally fed to SVM for depression classification. 
%
Instead of directly using original face videos, the baselines of AVEC 2016 and AVEC 2017 challenges \cite{valstar2016avec,ringeval2017avec} develop their ADA approaches based on audio behaviours and anonymous frame-level facial primitives to avoid ethical/privacy issues (discussed in Sec. \ref{subsec:facial behaviour ADA}), where hand-crafted prosodic, voice quality and spectral features are extracted to summarise short-term audio behaviours. 
Then, depression severity is predicted frame-wisely using RF regressor, and the unimodal clip-level audio and visual depression predictions are individually achieved by averaging all frame-level predictions, which are further averaged to achieve clip-level multi-modal ADA. Alghowinem et al. \cite{alghowinem2016multimodal} compute clip-level statistical features from not only audio behaviours but also frame-level eye activity and head poses, which are concatenated and undergo various feature selection processes, with the selected feature being fed to SVM for depression classification.

\textbf{Deep learning approaches:} A common pipeline for DL-based audio-visual ADA starts with learning short-term/frame-level behaviours, which are then encoded as clip-level representations to make the clip-level multi-modal depression assessment. For example, 
Uddin et al. \cite{uddin2022deep} first split the audio-visual data into a set of fixed-length segments and then feed them to a spatio-temporal network for depression-related segment-level multi-modal feature learning, identifying behaviours that are mostly contributed to depression assessment. These segment-level features are finally aggregated as clip-level representation for ADA. Tao et al. \cite{tao2024depmstat} separately encode the audio and visual segment-level features as clip-level representations via a temporal attentive pooling, which are then fused via a multi-modal factorised bilinear pooling to provide the final prediction. Niu et al. \cite{niu2020multimodal} propose a spatio-temporal attention network to separately learn short-term facial and audio depression cues, and then apply the eigen-evolution pooling to aggregate all short-term audio-visual cues as the clip-level representation. Finally, an attention module and a SVR complementarily combines  clip-level multi-modal representations for ADA.
Similarly, Yoon et al. \cite{yoon2022d} employ a two-stream transformer to learn clip-level facial and audio depression representations from all frame-level facial landmarks and audio LLDs, respectively. Then, a cross-attention module is adopted to capture the relationship between audio and visual features to learn a clip-level multi-modal representation for ADA. 
Focusing on fair ADA, Cheong et al. \cite{cheong2024fairrefuse} propose a causal multi-modal framework for debiasing multi-modal behaviours expressed by individuals with different backgrounds 
where audio and visual features are adaptively fused via a novel referee-based individual fairness guided fusion mechanism to make the multi-modal depression prediction.

\textbf{Hybrid approaches:} Some approaches also manage to leverage both domain knowledge provided by hand-crafted features and task-specific feature learning capability of the DL models for ADA. The AVEC 2019 challenge baseline \cite{ringeval2019avec} summarise various audio LLDs and facial behaviour primitives (e.g., AUs) of the given audio-visual clip. 
Jan et al. \cite{jan2017artificial} extract multiple hand-crafted frame-level facial features (LBP, EOH and LPQ), with a CNN learning depression-specific frame-level facial features. Meanwhile, spectral audio LLDs and MFCC features are extracted from every audio segment. The dynamics of these audio and visual features are then modelled by the feature dynamic history histogram (FDHH) for clip-level ADA. Hu et al. \cite{hu2024parallel} propose a two-stream CNN-LSTM network which separately capture short-term dynamics/features from all frame-level facial primitives (i.e., facial landmarks, gaze and AUs), and audio log-mel spectrograms. Then, a spatio-temporal attention pooling is proposed to obtain long-term audio-visual representations, which are fused at multiple temporal scales via a bridge fusion to output the clip-level multi-modal depression prediction.

\subsubsection{Audio-visual-language approaches}


\noindent \textbf{Hand-crafted approaches:} Based on audio, visual and texts, Williamson et al. \cite{williamson2016detecting} analyse both semantic contents and context from dialogue transcripts, where contents are modeled at the sentence-level via the GloVe embedding \cite{pennington2014glove} followed by PCA/ZCA to reduce dimension and sparse coding to extract final semantic features. To capture the joint dynamical facial behaviour properties during speech, a rank-ordered eigenspectrum is computed from multiple AUs, describing within-and cross-channel distributional properties of the multi-channel AU time series. Finally, the extracted audio (discussed in Sec. \ref{subsec:audio behaviour ADA}), visual and language features are individually fed to a Gaussian Staircase Model to make unimodal depression prediction, followed by a decision-level fusion. Building on a similar pipeline as AVEC 2013/2014 baselines, Pampouchidou et al. \cite{pampouchidou2016depression} achieve the clip-level multi-modal depression assessment via two strategies: (i) decision-level fusion of depression predictions made by clip-level visual representation (i.e., statistical and spectral features of facial landmarks, head motions and blinking rates), clip-level audio representation computed from various audio LLDs, and the clip-level language representation; and (ii) fusing multi-modal clip-level feature and then feeding it to a SVR or RF for ADA.
Gupta et al. \cite{gupta2014multimodal} extract frame-level LBP features and facial landmarks, alongside LBP-TOP and optical-flow-based motion vectors to capture short-term facial dynamics, all of which summarised as clip-level statistics. Meanwhile, multiple audio LLDs, spectral shape, spectro-temporal modulations, pitch periodicity, and long-term spectral variability, are jointly form a clip-level audio representation. Linguistic features are derived by mapping word sequences to affective ratings, processed via a Distributional Semantics Model and PCA as clip-level representation. The three modalities are then integrated and fed into an SVR for clip-level ADA. Nasir et al. \cite{nasir2016multimodal} compute  arithmetic means of: (1) frame-level pitch, glottal and voice quality-related as well as vocal tract audio (e.g., F1, F2 and F3) features; and (2) several facial behaviour primitives expressed in each 10s window. Additionally facial feature dynamics (i.e., velocity and acceleration), facial geometric and facial topology changes defined by facial landmarks are also computed. Finally, the feature-level fusion is achieved on clip-level audio, visual and language (statistics of phoneme rate and duration) representations, followed by RF for ADA. 


Alternatively, some multi-modal approaches learn visual depression cues from pre-extracted facial/visual behaviour primitives rather than original videos. Based on virtual assistant-guided sessions, Wu et al. \cite{wu2024mobile} separately conduct naive decision-level fusion and feature-level fusion for various mid-level and low-level hand-crafted features extracted from text, audio, facial attributes, heart rate, and eye movement, where MLP is employed as the classifier for four-level depression classification. 
Yang et al. \cite{yang2018integrating,yang2016decision} first select several equal-length video segment within every video to balance the depressed/non-depressed training examples. Based on each segment, it learns a Histogram of Displacement Range representing the range and speed of facial landmark displacements. Meanwhile, segment-level statistics summarising hand-crafted LLD audio features are computed. Finally, segment-level audio-visual features are processed by a two-stream CNN to individually make depression predictions, which are fused with textual features via a decision tree. Gong et al. \cite{gong2017topic} specifically explore the relationship between human audio, facial and language behaviours triggered by $83$ different interview topics defined by DAIC-WOZ database \cite{gratch2014distress} and the corresponding depression severity, where language information are partially used for topic modelling. For each topic, clip-level statistics of frame-level audio LLDs and facial descriptors as well as LIWC-based textual features, are extracted. Finally, a vector summarising all topic-level multi-modal features is constructed to make the clip-level depression prediction.

\textbf{Deep learning approaches:} CNN-RNN framework have been widely extended for audio-visual-language ADA. Zhang et al. \cite{zhang2024novel} deep learn frame-level/short-term audio, facial and language features via three CNNs, and then model their long-term dynamics via a BiLSTM. Then, the features learned from three modalities are fused via an attention, followed by a MLP to make the depression prediction. Wei et al. \cite{wei2022multi} first apply CNN-BiLSTM to independently learn audio, visual and textual features, and then introduce a novel sub-attention to fuse these multi-modal clip-level features. Lin et al. \cite{yin2019multi} also propose a Hierarchy BiLSTM to process audio, visual and text data for ADA. Rodrigues et al. \cite{rodrigues2019multimodal} apply a GNN-LSTM to process deep spectrum audio features and deep face features provided by ResNet-50. It additionally employs BERT to extract language features from texts, followed by a CNN-LSTM model to learn clip-level language depression features. Finally, multi-modal features are concatenated and fed to a dense layer for ADA. The similar strategy was also employed by \cite{ray2019multi}, where three BiLSTM networks are developed to obtain clip-level audio, visual and language representations from frame/segment-level hand-crafted and deep-learned features, followed by a multi-layer attention network to fuse them for ADA.
In \cite{zhang2024multimodal}, a ResNet and Swin Transformer-based dual visual encoder first learns global and local spatial facial features from each frame, and weights each facial local region's importance. A BiLSTM and a multi-head attention operation then model spatio-temporal depression cues from all frame-level representations. Meanwhile, the Mel spectrogram is calculated from the audio signal, followed by GRU and multi-head attention for temporal modelling. The text feature is modelled via a BERT Chinese model and attention. Finally, another attention module selects and combines depression features of three modalities, and a MLP is employed for the final prediction. 
Alternatively, some approaches achieve deep audio-visual-language ADA without using RNN-style networks, Haque et al. \cite{haque2018measuring} extend the Causal CNN to learn sentence-level embeddings from audio, text and 3D facial landmarks, which are summarised into a single vector via simple average at the sentence-level, and then fed to SVM for ADA. Sun et al. \cite{sun2022cubemlp} propose two audio, visual and language fusion strategies for ADA, where one forms an cubeMLP based on three independent MLP units to mix multi-modal features across three axes, while the other  \cite{sun2022tensorformer} designing a tensor-based multi-modal Transformer conducting source-target attentions between features of every modality pair. Zhang et al. \cite{zhang2024mddr} introduces two attention-based aggregation strategies to: (1) dynamically evaluate the relevance among audio, visual and language features to adaptively integrate their temporal dynamics; and (2) determine the importance of frame-level amalgamated features to select the most depression-specific features. To investigate dynamics of multi-modal behaviours, Shen et al. \cite{shen2024multi} propose a multi-modal behaviour graph representation learning framework which learns a set of heterogeneous node features to represent short segment-level facial behaviours, long segment-level audio behaviours, and the global text cues. Then, edges are constructed to connect nodes of the same and different modalities, to model multi-scale intra and inter-modal dynamics. This clip-level graph representation is then fed to a GNN for ADA. 

\subsubsection{Other multi-modal approaches}

\noindent A few studies also utilise other modality combination for ADA. Liu et al \cite{liu2022measuring} propose a two-stream CNN for facial and body movement-based ADA, where one stream is pre-trained to predict frame-level facial expression, based on which a video-level expression entropy is computed. The other stream locates the position of the individual's 18 body joints and calculate their video-level motion amplitude. Finally, a LR fuses video-level facial expression entropy and body joint motion amplitude for ADA. Ning et al. \cite{ning2024depression} propose an auxiliary decision-making system to combine EEG and speech signals for ADA. It explores both linear and nonlinear features from each modality, and concatenates them for depression classification. Li et al. \cite{li2024sparse} fuse a sparse emotion dictionary describing textual data and spectral audio features for ADA, where a multi-head self-attention-based low and high frequency feature fusion strategy is proposed. Ye et al. \cite{ye2021multi} extract several hand-crafted audio LLDs and apply DeepSpectrum to deep learn depression-related audio features, which are subsequently integrated into a customised TCN, with the final layer employing relational attention classification. Meanwhile, they employ the Continuous Bag of Words (CBOW) method for text feature extraction. Both clip-level features are concatenated and fed to a MLP for ADA. 
Zheng et al. \cite{zheng2023two} propose a temporal convolution transformer to jointly assess depression and emotion status from human audio, visual, text and additional physiological behaviours. Fan et al. \cite{fan2024transformer} leverage CNN, transformer and GNN to extract depression-related frame/segment-level video, audio and remote photoplethysmographic (rPPG) features and treats each of them as a graph node. Then, a transformer is introduced to model inter-modal, intra-modal, and tri-modal relationships among nodes learned from three modalities. Finally, a GNN fuses all features for clip-level multi-modal ADA.
Lin et al. \cite{lin2021looking} extract various hand-crafted frame-level features from human body movements, facial behaviours and audio signals, which are summarised as a multi-modal clip-level representation through a Multi-modal Deep Denoising Auto-Encoder and Fisher vector encoding, followed by a RF for feature selection. Finally, MLP and LR are separately applied for ADA. Shao et al. \cite{shao2024multimodal} propose to learn complementary depression-related cues from brain behaviours detected by EEG and fNIRS, where spatio-temporal CNNs and transformers are stacked to jointly process both modalities.

\subsection{Statistics and trend}

\begin{figure*}
\centering
\includegraphics[width=13.9cm]{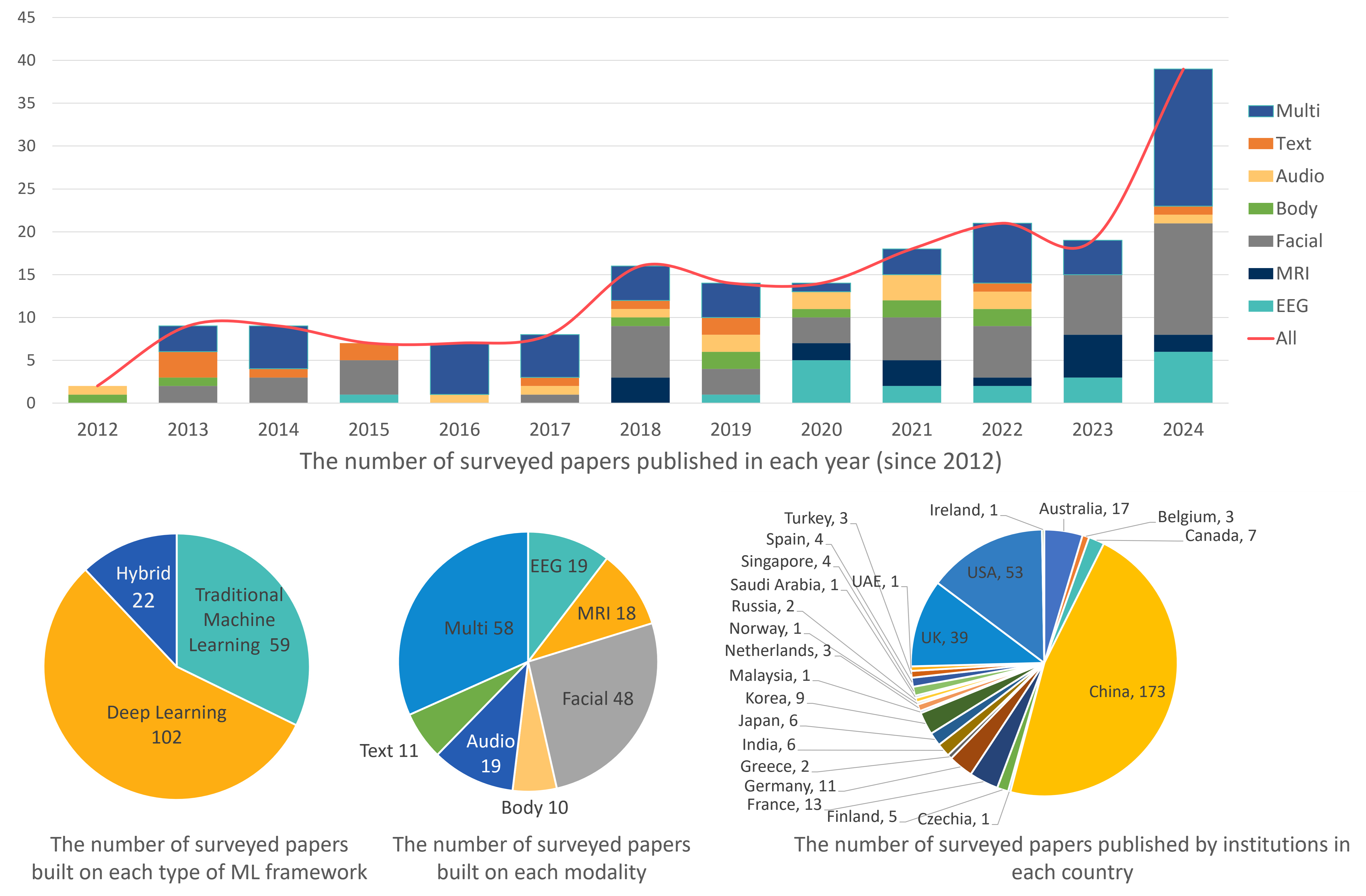}
\caption{Statistics and trend of ADA approaches (2012-2024) surveyed in this paper.}
\label{fig:statistics_trend}
\end{figure*}

\noindent Fig. \ref{fig:statistics_trend} illustrates statistics computed from ADA publications considered by this survey. It can be observed that the number of ADA publications, especially for face, text and multi-modal based approaches, has drawn clear attentions since 2013. This may largely contributed by the AVEC 2013 and AVEC 2014 challenges that provided initial venues and open-source audio-visual ADA datasets for developing and comparing audio-visual ADA solutions.  
The small drop in 2019 might caused by the fact that there was no depression sub-challenge in AVEC 2018. Importantly, the number of ADA publications has been largely increased since 2018, suggesting that the ADA research has drawn more attentions in recent years, even during and after the COVID pandemic. Among all surveyed ADA solutions (since 2012), deep learning models have been applied to more than half of them. From another perspective, the most popular uni-modality for the surveyed ADA methods is the facial behaviour. Although a large number of audio and text-based depression feature learning strategies have been investigated, they were generally combined or even integrated with the visual modality to form multi-modal ADA methods. Authors affiliated with institutions in China, USA, and UK developed most of existing ADA solutions, with the remainder mainly coming from Institutions of other Europe and Asia countries (i.e., only counting first and last authors of each publication).

\section{Datasets and Competitions}
\label{sec: dataset and challenge}

\subsection{Publicly available ADA datasets}



\noindent This section reviews existing \textit{publicly available} ADA datasets. The \textbf{main challenge is their limited scale in participant numbers and clip lengths}, i.e., they contain (except the V-log dataset that lacks rigorous labeling) fewer than 200 participants with each clip usually lasting less than 30 mins (i.e., depression is a long-term condition reflected over extended periods). \textbf{Another limitation is that depression labels are typically obtained from self-reported questionnaires rather than standardised clinical assessments}, leading to some degrees of subjectivity and inconsistency.  Such issues not only limit the advancement of DL-based ADA models but also undermine evaluation objectivity. 

The \textbf{AVEC 2013 depression corpus} \cite{valstar2013avec} is a pioneering publicly available audio-visual ADA dataset containing 150 naturalistic clips of individuals engaging in various human-computer interaction (HCI) tasks guided by slides, recorded via a laptop camera and microphone. 
Its subset, the \textbf{AVEC 2014 depression corpus} \cite{valstar2014avec}, only includes two HCI tasks: (1) FreeForm, where participants respond to predefined questions in German; and (2) NorthWind, where they read an excerpt from Die Sonne und der Wind. Unlike AVEC 2013, each task is treated as an independent clip, yielding 300 short clips. Both datasets provide clip-level depression severity annotations based on BDI-II scores. However, a key limitation is the overlap of individuals between training/validation and test sets, making the evaluation  less objective. While the AVEC 2013/2014 datasets are limited in size for training and evaluation, the \textbf{Vlog Depression Dataset} \cite{yoon2022d} offers a large-scale alternative. It contains 961 vlogs from 816 individuals with a balanced distribution of depressed and non-depressed examples. Unlike the above controlled datasets, these vlogs were obtained from YouTube, where individuals exhibit natural, unrestricted behaviours. However, their binary depression labels were assigned solely by non-expert annotators (college students) based on vlog content and titles, which introduces potential label reliability concerns.

Alternatively, a large part of public ADA datasets were collected under clinical interview-style scenarios. The \textbf{DAIC-WOZ dataset} \cite{gratch2014distress} is an audio-visual-text corpus recorded under several semi-structured interviews designed to assess PTSD and depression. Participants undergo one of four interview types: face-to-face, video conference, Wizard-of-Oz (human-controlled virtual interviewer), or pre-defined virtual agent-controlled interviews. To avoid ethical risks, the dataset provides de-identified audio, text transcripts, and mid-/low-level anonymous facial behaviour features extracted by OpenFace \cite{baltruvsaitis2016openface}, rather than raw facial videos. Clip-level depression severity is annotated using self-reported PHQ-8 scores (PHQ-9 without the suicide question). Its extension, \textbf{Extended DAIC-WOZ} \cite{ringeval2019avec}, used for the AVEC 2019 challenge, follows the same protocol but includes more clips (i.e., 275 clips from 275 individuals), which incorporates additional deep spectrum audio features and deep facial features extracted via VGG-16 and ResNet-50. Similarly, the \textbf{Chinese Multimodal Depression Corpus (CMDC)} \cite{zou2022semi} also comprises face-to-face human interviews conducted under strict participant selection criteria, with assessments by professional psychiatrists. Interviews follow pre-defined topics based on clinical practice and depression questionnaires, yielding de-identified transcripts, audio signals, and OpenFace facial primitives, where clip-level depression labels are defined by HAMD-17 and PHQ-9 scores. The \textbf{Pittsburgh Clinical Depression (PCD) dataset} \cite{dibekliouglu2018dynamic} also records in-person face-to-face clinical interviews with participants undergoing treatment via either selective serotonin re-uptake inhibitor (SSRI) or Interpersonal Psychotherapy (IPT), where interview questions and depression categories are both defined by the HAM-D questionnaire.

In addition to the above datasets recording facial, audio and/or textual behaviours, other datasets explore alternative modalities for ADA. The \textbf{Looking at the Body (LaB) dataset} \cite{lin2021looking} captures full-body motions, gestures, facial behaviours and speeches from 35 gender-balanced participants in face-to-face interviews with open-ended questions. Each of its clip is annotated with self-reported PHQ-8 scores, along with anxiety, stress, somatic symptom burden, and personality labels. 
The \textbf{Emotional Audio-Textual Depression Corpus (EATD-Corpus)} \cite{shen2022automatic} contains audio and transcript recordings from 162 Chinese students participating in APP-based online interviews. Each participant answered three randomly selected questions, with depression severity labeled using the self-reported SDS questionnaire \cite{zung1965self}. 
The \textbf{MODMA dataset} \cite{cai2022multi} record EEG and speech data from 24 clinically depressed patients and 29 healthy individuals. It includes: (1) 128-electrode EEG recordings during resting state and a Dot Probe task; (2) 3-electrode EEG signals in resting state; and (3) audio recordings from tasks of 18-question interviews, word-reading, and facial expression picture description. The binary depression labels were manually assigned by professional psychiatrists.
Although the lack of large-scale and well-annotated dataset challenge remains, the above datasets form a first step towards addressing this gap.
Overall, these are the main datasets made publicly available and have been more commonly researched within the community. 

\subsection{Grand challenges (Competitions)}




To provide venues for researchers showing their ADA solutions and exchanging ideas, a series of AVEC depression challenges have been organized, which mainly focused on inferring depression via non-invasive ways, i.e., through non-verbal audio-visual spatio-temporal behaviours and verbal textual cues. Specifically, \textbf{the AVEC 2013 depression challenge} \cite{valstar2013avec} is the first public audio-visual ADA challenge facilitating researchers from various institutions to extend the analysis of affective behaviours to infer a more complex mental state: depression severity, which brought together the initial efforts for audio-visual ADA. Its subsequent \textbf{AVEC 2014 challenge} \cite{valstar2014avec} follows the same manner but only used two task-dependent subsets (i.e., AVEC 2014 depression dataset). In both challenges, participant teams were required to develop their models based on original face videos and German speech signals. The ADA baselines of both challenges extract various hand-crafted audio and visual features while leveraging several traditional ML regressors for clip-level depression severity estimation (please check the main paper for details). The Root Mean Square Error (RMSE) and Mean Absolute Error (MAE) were employed as metrics for the evaluation.




The following AVEC 2016 \cite{valstar2016avec}, AVEC 2017 \cite{ringeval2017avec} and the most recent AVEC 2019 depression \cite{ringeval2019avec} challenges focused on de-identified multi-modal human behaviour-based ADA, which additionally introduced verbal text modality. They employed DAIZ-WOZ depression dataset \cite{gratch2014distress} or its extension to provide a common ADA benchmark under well-defined and strictly comparable semi-structured clinical interview-based conditions. In these challenges, only de-identified audio and text data as well as anonymous facial behaviour primitives/attributes (e.g., AUs, facial landmarks, head poses, gazes, HOG features and discrete emotions) were provided. Participant teams were required to develop their ADA system based on such de-identified behaviour data rather than original audio signals and face videos. Besides the depression severity estimation, these challenges also provide baselines for binary depression classification. Again, the RMSE and MAE were used for evaluating depression severity estimation systems, while F1, Precision and Recall were adopted to evaluate the depression classification systems. Additionally, the Concordance Correlation Coefficient (CCC) that takes both Mean Square Error (MSE) and Pearson Correlation Coefficient (PCC) between depression severity predictions and their ground-truths into consideration was used to rank all participant teams in AVEC 2019.

\section{Conclusion and Discussion}
\label{sec:discuss}

\noindent This paper conducted the first comprehensive survey reviewing various human behaviour-based automatic depression assessment (ADA) approaches, covering both internal brain activities (recorded by MRI and EEG) as well as expressive facial, body, audio and language behaviours. We also discuss depression-related human internal and external behaviours as well as  publicly available ADA datasets in details, providing theoretical and practical basis for the surveyed ADA approaches.
We present our findings of 183 ADA papers and ten datasets at a centralised web repository \footnote{\url{https://github.com/SSYSteve/Automatic-Depression-Assessment-Survey}} to facilitate the dissemination of our findings.
We call for concerted efforts towards the creation of a common repository of resources to support researchers navigating the ADA field. 
Inspired by the initiatives such as the 
Physionet database/HumaneAI network \footnote{https://physionet.org/ ,  https://www.humane-ai.eu/}, we suggest that the \textit{depression detection common repository} should include:
(i) a collection of depression-detection dataset and its collection procedure;
(ii) a library of commonly used libraries, tools and benchmark models to reproduce existing SOTA results; and 
(iii) experimental guidelines/frameworks to ensure that research conducted adheres to ethical guidelines. 

Moreover, given the innately private and sensitive nature of ADA research, having a centralised repository of resources may introduce new ethical challenges and concerns. 
Thus, we propose to cohesively address these issues together as a community. 
These can be done via leveraging more ethically informed frameworks, further engagement with the different communities and social groups, making research more transparent and open-source and encouraging deeper cooperation between stakeholders.
Our review seeks to provide the different parties with the necessary overview of the field to facilitate such cooperation.
%

\subsection{Challenges}

As our findings above show, although there is a significant amount of effort channeled into advancing the field of ADA, there are several key factors hampering significant development and adoption in this area.
Some of the key 
challenges of existing ADA research include ML bias in ADA models, lack of interpretability and reproducibility, lack of generalisable solutions, as well as the lack of large-scale and high-quality datasets.
%
%

\textbf{ML bias in ADA:} This is a growing source of concern \cite{cheong2023towards,bias_survey}, including variability in depression manifestation \cite{cheong2023towards,cameron2024multimodal} and the small dataset challenge \cite{small_but_fair}. Although some works have made progress on this front \cite{cameron2024multimodal,cheong2024fairrefuse,cheong_ufair_pmlr}, there is still a need to develop principled methods to account for other sources of bias such as labeling bias \cite{bias_survey} and address broader notions of fairness beyond parity-based fairness measures \cite{barocas2023fairness, counterfactual_fairness,kuzucu2024uncertainty,bias_survey}. Moreover, no existing work has investigated the interpretability, explainability, nor reproducibility of such models.


\textbf{Lack of interpretability and reproducibility:} 
Most advanced DL-based ADA approaches end-to-end infer depression predictions from human behaviours. Despite their effectiveness, these `black-box' models lack explicit rule-based logic explaining how they extract depression-related cues and make predictions. Specifically, they fail to clarify: (i) which behavioural patterns and their combinations are most predictive of depression, (ii) the complementary cues provided by different modalities; and (iii) why certain models outperform others in identifying depression-related behaviours. Furthermore, most of them have limited code availability, making them difficult to be fully re-produced, i.e., only a few of them have fully opensourced their implementations \cite{song2022spectral,gimeno2024reading,xu2024two,islam2024facepsy,pan2024spatial,shen2022automatic}.

\textbf{Lack of generic solutions to tackle with common data issues:} Although some approaches \cite{song2022spectral,xu2024two} attempted to learn video-level fixed-size representations, DL-based ADA models typically struggle with processing long and variable-length face videos without significant information lose/distortion, due to DL models' fixed-size input requirements and high computational costs of processing all frames. Second, there lacks standardised and solid solutions to handle noises within depression data, e.g., low signal-to-noise EEG, speech interference, or missing target faces in video frames, despite some efforts have been already made \cite{song2024loss,he2022reducing}. Third, cultural and linguistic diversity limits the generalizability of language-based ADA models \cite{ye2021multi,trotzek2018utilizing,cameron2024multimodal}, resulting in poor performances under different linguistic and cultural contexts. 



\textbf{Lack of large-scale and high-quality ADA datasets:} Another vital challenge is the lack of large-scale, high-quality and well-annotated publicly available datasets for researchers to develop and fairly compare their expressive behaviour-based ADA models. 
Depression labels are typically obtained from self-reported questionnaires rather than standardised clinical assessments
%
which not only limit the advancement of DL-based ADA models but also undermine evaluation objectivity (please refer to Sec. \ref{sec: dataset and challenge} for more details). 
Alternatively, collecting a high-quality and large-scale depression dataset with diverse participant demographics and reliable depression labels remains challenging due to ethical, financial, data anonymisation, data collection resource and high-quality clinical labeling constraints.

\subsection{Opportunities and future directions}

\noindent 
There are multiple avenues for enhancing DL-based ADA solutions. 
First, researchers could start adding a bias analysis section in their ADA publications (e.g., similarly to the recommendations made within the affective computing community \cite{cheong_acii}). 
%
%
%
Second, further research could focus on developing hand-crafted behavioural descriptors for depression, grounded in physiological and psychological insights. Such descriptors would not only enhance interpretability but also guide the extraction of meaningful features for depression detection.
%
%
Third, researchers can develop generalizable strategies to identify which input behaviours contribute most significantly to ADA, and which behaviours do not effectively distinguish between depressed and non-depressed individuals.
%
%
%
%
%
Fourth, researchers could explore simulating an individual’s internal, personalized cognition based on their expressive behaviours to infer depression status \-- an approach inspired by similar strategies used in true personality recognition tasks \cite{song2022learning,song2021self}. Fifth, future work could focus on reliably eliciting depression-related behaviours in an unobtrusive manner to enable more effective and natural data collection. Finally, establishing a research community around the ADA problem encouraging open science would facilitate fair comparisons, reproducibility, and broader collaboration. In summary, our work aims to contribute to the ADA field by offering a comprehensive overview, identifying current limitations, and outlining opportunities to foster cohesive collaboration and promote more ethical and impactful ADA research.


%




\ifCLASSOPTIONcaptionsoff
  \newpage
\fi



%



\bibliographystyle{IEEEtran}

\bibliography{egbib}

\end{document}